\documentclass{elsarticle} 

\usepackage{fullpage}
\usepackage{amsmath} 
\allowdisplaybreaks 
\usepackage{multirow} 
\newcommand{\astc}{\textrm{\hspace{1pt}\raisebox{1pt}{$\ast$}}} 

\journal{arxiv.org}

\biboptions{authoryear}

\begin{document}

\begin{frontmatter}

\title{The Consequences of Switching Strategies \\ in a Two-Player Iterated Survival Game}

\author[ens]{Olivier Salagnac}
\ead{olivier.salagnac@ens.fr}
\author[oeb]{John Wakeley}
\ead{wakeley@fas.harvard.edu}

\address[ens]{\'{E}cole Normale Sup\'{e}rieure, Paris, Cedex 05, France}
\address[oeb]{Department of Organismic and Evolutionary Biology, Harvard University, Cambridge, MA, 02138, USA}

\begin{abstract}
We consider two-player iterated survival games in which players may switch from a more cooperative behavior to a less cooperative one at some step of the game.  Payoffs are survival probabilities and lone individuals have to finish the game on their own.  We explore the potential of these games to support cooperation, focusing on the case in which each single step is a Prisoner's Dilemma.  We find that incentives for or against cooperation depend on the number of defections at the end of the game, as opposed to the number of steps in the game.  Broadly, cooperation is supported when the survival prospects of lone individuals are relatively bleak.  Specifically, we find three critical values or cutoffs for the loner survival probability which, in concert with other survival parameters, determine the incentives for or against cooperation.  One cutoff determines the existence of an optimal number of defections against a fully cooperative partner, one determines whether additional defections eventually become disfavored as the number of defections by the partner increases, and one determines whether additional cooperations eventually become favored as the number of defections by the partner increases.  We obtain expressions for these switch-points and for optimal numbers of defections against partners with various strategies.  These typically involve small numbers of defections even in very long games. We show that potentially long stretches of equilibria may exist, in which there is no incentive to defect more or cooperate more.  We describe how individuals find equilibria in best-response walks among strategies, and establish that evolutionary stability requires there be just one such equilibrium.  Otherwise, equilibria are not protected against invasion by strategies with fewer defections.   
\end{abstract}

\end{frontmatter}


\section{Introduction} \label{sec:intro}

In a two-player iterated survival game, individuals may or may not survive each step and an individual whose partner has died must continue alone \citep{EshelAndWeinshall1988}.  It is a game against Nature \citep{Lewontin1961} such as when individuals have to fend off repeated attacks by a predator \citep{Garay2009,DeJaegherAndHoyer2016} or face other sorts of adversity \citep{Emlen1982,Harms2001,SmaldinoEtAl2013,DeJaegher2019}.  These may include harsh physical conditions.  As \citet[p.\ 69]{Darwin1859} had noted: ``When we reach the Arctic regions, or snow-capped summits, or absolute deserts, the struggle for life is almost exclusively with the elements.''  Observing animals living together under harsh physical and biological conditions, \citet{Kropotkin1902} suggested that mutual aid is all but inevitable in evolution.  Iterated survival games are a simple way to model these scenarios, and they do show that, when the prospects for lone individuals are not great, self-sacrificing cooperative behaviors can be strongly favored \citep{EshelAndWeinshall1988,EshelAndShaked2001,Garay2009,WakeleyAndNowak2019}.  

We consider iterated survival games of fixed length $n$.  We assume that there are two possible single-step strategies or behaviors, which we call $C$ and $D$.  The probability an individual lives through a single step is given by Table~\ref{tab:twobytwo}, and the game is symmetric in the sense that both players receive payoffs (live or die in each step) according to this matrix.  The choice of labels $C$ and $D$ coincides with an assumption, $a>d$, that individuals in $CC$ pairs fare better than individuals in $DD$ pairs.  Total payoffs, which are overall survival probabilities, accrue multiplicatively across the $n$ steps.  These depend on the overall strategies of individuals, which are fixed strings of $C$s and $D$s.  We ask whether it might be advantageous to switch from a more cooperative behavior ($C$) to a less cooperative behavior ($D$) at some step of the game.  

\newlength{\tempdima}
\settowidth{\tempdima}{Partner}
\addtolength{\tempdima}{-2\tabcolsep}
\begin{table}[h]
 \hspace{1.75in}
 \begin{tabular}{ccccc}
  & & \multicolumn{3}{c}{Partner} \\
  & & {\makebox[0.5\tempdima]{$C$}} & {\makebox[0.5\tempdima]{$D$}} & {\makebox[0.5\tempdima]{$\O$}}  \\ 
  \cline{3-5}
  \multirow{2}{*}{Individual} & $C$ & \multicolumn{1}{|c}{$a$} & \multicolumn{1}{|c|}{$b$} & \multicolumn{1}{|c|}{$a_0$} \\
  \cline{3-5}
  & $D$ & \multicolumn{1}{|c}{$c$} & \multicolumn{1}{|c|}{$d$} & \multicolumn{1}{|c|}{$a_0$}\\
  \cline{3-5}
 \end{tabular}
\caption{The single-step payoff ($a$, $b$, $c$, $d$ or $a_0$) in a symmetric two-player survival game is the probability of survival of an individual when the individual and partner have specified single-step strategies, either $C$ or $D$, or when the individual is playing alone because the partner has died $(\O)$.  The loner survival probability, $a_0$, does not depend on the individual's strategy.  }  \label{tab:twobytwo}
\end{table}

From the standpoint of behavioral biology or mathematical ecology, this is a phenomenological rather than a mechanistic model \citep{GeritzAndKisdi2012}.  It is described plainly in terms of the relative survival of types in different combinations, and skirts any details about `who helps whom achieve what' \citep{RodriguesAndKokko2016}.  Survival is an obviously crucial kind of utility for individuals, which also combines in various ways with fertility to produce evolutionary fitness \citep{ArgasinskiAndBroom2013}.  Here, when we address evolutionary stability, we assume there are no differences in fertility.  The principal assumptions we make are that the single-step payoffs ($a,b,c,d,a_0$) are fixed for the entire game, and that survival outcomes are statistically independent both in different steps and for different players in a single step.  The consequent multiplicative accrual of payoffs turns relatively mild single-step games into mortally challenging iterated games as $n$ increases.  This naturally produces strong interdependence between individuals, which is known to favor cooperation and is purposely assumed in other models \citep{Roberts2004}.

When both players are present, then depending on the magnitudes of $a$ versus $c$ and $b$ versus $d$, each step will fall into one of the four well-known classes of symmetric two-player games.  Ignoring the possibility that some payoffs might be equal: $a<c$ and $b<d$ defines the class of games represented by the Prisoner's Dilemma \citep{Tucker1950,RapoportAndChammah1965}; $a>c$ and $b<d$ defines the class represented by the Stag Hunt \citep{Skyrms2004}; $a<c$ and $b>d$ defines the class represented by the Hawk-Dove game \citep{MaynardSmithAndPrice1973,MaynardSmith1978}; and $a>c$ and $b>d$ defines the class which was recently dubbed the Harmony Game \citep{DeJaegherAndHoyer2016}.  In the case of the Prisoner's Dilemma, $a$ corresponds to the ``reward'' payoff, $b$ to the ``sucker's'' payoff, $c$  to the ``temptation'' payoff, and $d$ to the ``punishment'' payoff \citep{RapoportAndChammah1965}.  
 
\citet{WakeleyAndNowak2019} considered individuals with constant strategies (all-$C$ or all-$D$) and studied how the relative frequency of the cooperative type changes over time in a well-mixed population due to differential death in the two-player iterated survival game.  Depending especially on the number of iterations $n$ and the loner survival probability $a_0$, the $n$-step game may be of a different type than the single-step game, with obvious implications for the evolution of cooperation.  For example, if $n$ is large and $a_0$ is small, the $n$-step game may be a Harmony Game even if the single-step game is a Prisoner's Dilemma.  Then cooperation is favored despite the fact that it seems better to defect in any given step.  On the other hand, if $a_0$ is large, the $n$-step game may favor all-$D$ even if the single-step game is a Harmony Game.    

Here we study the problem of optimal strategy choice for a broader range of $n$-step strategies, specifically ones which switch from $C$ to $D$ at some step of the game.  Strategy $S_i$ plays $D$ for the final $i$ steps of the game (and $C$ for the first $n-i$ steps) where $i$ can range from $0$ to $n$.  Thus, $S_0$ is all-$C$ and $S_n$ is all-$D$.  We study the general case of an $S_j$ individual with an $S_i$ partner, and ask whether there is an advantage to increasing or decreasing $j$ depending on the other six parameters $(a,b,c,d,a_0,n)$.  We are interested in the presence of optima, for which there is no incentive for the individual to increase or decrease the number of defections.  We find critical values of $a_0$ which determine the strategy choice of individuals.  Broadly, if $a_0$ is large, then all-$D$ is the only optimum, whereas if $a_0$ is small, then a single intermediate optimum or a stretch of intermediate optima may exist.  For moderate $a_0$, is it also possible that no strategies are optimal, that instead incentives exist both to increase and to decrease the number of defections.  

We focus primarily on the case where the single step game is a Prisoner's Dilemma.  Comprehensive treatment of this case uncovers an unexpected array of possible behaviors.  With reference to questions about the incentives for cooperation, our results illustrate that when individuals depend very strongly on their partners, the motivation to defect or otherwise be non-cooperative may be dramatically less than is typically understood from the analysis of standard models of repeated games with additive payoffs.

\section{Markov model of individual survival and preliminary calculations} \label{sec:markov}

The survival game is symmetric, so we can focus on one player, nominally the individual of Table~\ref{tab:twobytwo}.  The individual is in one of three possible situations: alive with a partner, alive without a partner or dead.  We use a Markov chain to model transitions among these three states.  The probabilities of surviving to the next round are given by Table~\ref{tab:twobytwo}, symmetrically for both players, and players live or die independently of one another other in each step of the game.  The chain is non-homogenous because transition probabilities depend on the strategies of the individual and the partner.  There are four possible pairs of single-step strategies for the individual (listed first) and the partner (listed second) when both are present---$CC$, $DC$, $CD$, and $DD$---and we use these to index four corresponding single-step transition matrices.  We use $\O$ to denote that one of the players has died and $\astc$ as a placeholder for the partner when the individual has died.  The game always starts with two players, but then changes state randomly according to these matrices.
\begin{align}
    \bordermatrix{ & CC & C\O & \O\astc \cr
    CC & a^2 & a(1-a) & 1-a \cr
    C\O & 0 & a_0 & 1-a_0 \cr
    \O\astc & 0 & 0 & 1} = A_{CC} \label{eq:acc}
    \\[6pt] 
     \bordermatrix{ & DC & D\O & \O\astc \cr
    DC & bc & c(1-b) & 1-c \cr
    D\O & 0 & a_0 & 1-a_0 \cr
    \O\astc & 0 & 0 & 1} = A_{DC} \label{eq:adc}
    \\[6pt]
     \bordermatrix{ & CD & C\O & \O\astc \cr
    CD & bc & b(1-c) & 1-b \cr
    C\O & 0 & a_0 & 1-a_0 \cr
    \O\astc & 0 & 0 & 1} = A_{CD} \label{eq:acd} 
    \\[6pt]
     \bordermatrix{ & DD & D\O & \O\astc \cr
    DD & d^2 & d(1-d) & 1-d \cr
    D\O & 0 & a_0 & 1-a_0 \cr
    \O\astc & 0 & 0 & 1} = A_{DD} \label{eq:add}
\end{align}

\begin{figure}[h]
\bigskip
\centering 
\includegraphics{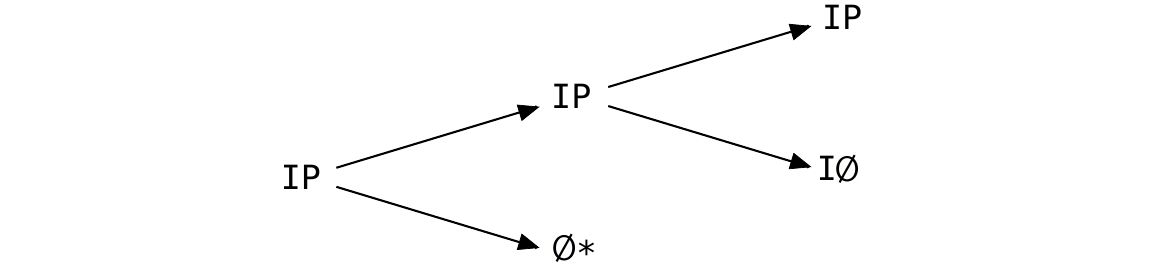}
\caption{Two-event decomposition of a single step in the iterated survival game when both players are present, illustrating Individual-Partner dependence.  The diagram can be used to compute the first-rows transition probabilities in the matrices in (\ref{eq:acc}) through~(\ref{eq:add}) by replacing {\small \texttt{I}} and {\small \texttt{P}} with strategies $C$ or $D$ then assigning probabilities to the arrows.} \label{fig:pairtree}  
\end{figure}

The second and third rows of all four matrices are identical due to our assumption of a single loner survival probability regardless of strategy, and because the state $\O\astc$ is absorbing for an individual.  The transitions described by the first rows of the matrices are more complex because they involve two events, one for the individual and one for the partner.  Although payoffs are awarded simultaneously to both players in determining the transition probabilities in the first rows, this two-fold structure lends itself to depiction as an extensive form of the single-step game between two players \citep{vonNeumann1928,Kuhn1953,Cressman2005}.  This is illustrated in Fig.~\ref{fig:pairtree} and underscores the strong dependence between players in an iterated survival game.  Figure~\ref{fig:pairtree} is also a probability tree diagram because the transition probabilities in the first rows in (\ref{eq:acc}) through (\ref{eq:add}) can be obtained by multiplying probabilities associated with the arrows. 

An individual with a partner may die, in which case the game is over for the individual regardless of what happens to the partner.  This event is represented by the first down-arrow in Fig.~\ref{fig:pairtree}.  Having a large survival probability when the partner is present is the only protection against this fate for the individual.  Here, the usual comparisons of $a$ versus $c$ and $b$ versus $d$ describe the consequence of switching strategies against a partner with a given strategy.  But the future state of the individual also depends on what happens to the partner.  If the partner dies (second down-arrow in Fig.~\ref{fig:pairtree}), the individual ends up alone and will be subject to the loner survival probability in every remaining step of the game.  

The only way to remain in state one of the Markov chain is for both players to survive (both up-arrows in Fig.~\ref{fig:pairtree}).  The probability of this combined event is given by the upper-left or (1,1) entries in each matrix, which depend on the strategies of both players.  Thus, the consequences of switching strategies will also depend on the comparisons of $a^2$ versus $bc$ and $bc$ versus $d^2$.  This can be understood in terms of the number of cooperators in each possible pair of single-step strategies.  Switching from $D$ to $C$ against a $D$ partner changes the number of cooperators in the pair from zero to one, and switching from $D$ to $C$ against a $C$ partner changes it from one to two.  The inclusion of the first cooperator in a pair has effect $bc-d^2$ whereas the inclusion of a second cooperator has effect $a^2-bc$.  Then, for example, an individual who suffers a cost $b-d<0$ in a Prisoner's Dilemma might also enjoy the benefit of not having to survive alone, if it is also true that $bc-d^2>0$. 

The series of single-step strategies in the game between an individual with $n$-step strategy $S^{(n)}_j$ and a partner with $n$-step strategy $S^{(n)}_i$, which we write simply as $S_j$ and $S_i$, may be depicted as  
\begin{equation}
\begin{aligned}
    S_j &= \underbrace{CCC\dots C}_{n-j ~ \mathrm{steps}} \underbrace{DDD\dots D}_{j-i~ \mathrm{steps}}\underbrace{DDD\dots D}_{i~ \mathrm{steps}}\\[-4pt]
    S_i &=\overbrace{CCC\dots C}_{} \overbrace{CCC\dots C}_{ }\overbrace{DDD\dots D}_{} 
\end{aligned} \label{eq:series}
\end{equation}   
for $i,j\in[0,n]$ and $j \geq i$.  Our goal is to understand the overall survival of $S_j$ when paired with $S_i$ for any given $(i,j)$.  Any such game can be partitioned into three phases: both players having strategy $C$, one $C$ and one $D$, and both $D$.  The ordered series of these will determine the overall transition matrix.  For the example in (\ref{eq:series}), we have the product $A^{n-j}_{CC} A^{j-i}_{DC} A_{DD}^{i}$. 

We employ the following decomposition---exemplified by the case $CC$, when both players having strategy $C$---in order to compute the powers of the four matrices. 
\begin{align}
    A_{CC} &= \begin{pmatrix}a^2 & a(1-a) & 1-a \\ 0 & a_0 & 1-a_0 \\ 0 & 0 & 1 \end{pmatrix} \notag \\[6pt] 
    &=\begin{pmatrix}1&a(1-a)&1\\0&a_0 -a^2 &1\\0&0&1\end{pmatrix} \begin{pmatrix}a^2 &0&0\\0&a_0&0\\0&0&1\end{pmatrix} \begin{pmatrix}1&\frac{a(a-1)}{a_0 - a^2} & \frac{a-a_0}{a_0 -a^2} \\ 0& \frac{1}{a_0 -a^2} & \frac{-1}{a_0 -a^2} \\ 0&0&1\end{pmatrix} . \label{eq:decompose}
\end{align}
The diagonal elements in the middle matrix in (\ref{eq:decompose}) and in $A_{CC}$ itself are the eigenvalues of $A_{CC}$.  The two outer matrices in (\ref{eq:decompose}) are the inverses of each other.  For any $k = 0,1,2,\ldots$, we have 
\begin{align}
    A_{CC}^k &= \begin{pmatrix}1&a(1-a)&1\\0&a_0 -a^2 &1\\0&0&1\end{pmatrix} 
    \begin{pmatrix} a^{2k} &0&0 \\ 0&a_0^k&0 \\ 0&0&1 \end{pmatrix} 
    \begin{pmatrix}1&\frac{a(a-1)}{a_0 - a^2} & \frac{a-a_0}{a_0 -a^2} \\ 0& \frac{1}{a_0 -a^2} & \frac{-1}{a_0 -a^2} \\ 0&0&1\end{pmatrix}\\[6pt]
    &=\begin{pmatrix}a^{2k}& a(1-a)\frac{a_0^k-a^{2k}}{a_0 -a^2} & 1-a^{2k}-a(1-a)\frac{a_0^k - a^{2k}}{a_0 - a^2} \\ 0 & a_0^k & 1-a_0^k \\ 0&0&1\end{pmatrix} . \label{eq:ACCk}
\end{align}
Applying the same technique to $A_{DC}$, $A_{CD}$ and $A_{DD}$ we obtain
\begin{align}
& A_{DC}^{k} =\begin{pmatrix}(bc)^{k}& c(1-b)\frac{a_0 ^{k}-(bc)^{k}}{a_0 -bc} & 1-(bc)^{k}-c(1-b)\frac{a_0 ^{k}-bc^{k}}{a_0 -bc} \\ 0 & a_0 ^{k} & 1-a_0 ^{k} \\ 0&0&1\end{pmatrix} \\[6pt]
& A_{CD}^{k} =\begin{pmatrix}(bc)^{k}& b(1-c)\frac{a_0 ^{k}-(bc)^{k}}{a_0 -bc} & 1-(bc)^{k}-b(1-c)\frac{a_0 ^{k}-bc^{k}}{a_0 -bc} \\ 0 & a_0 ^{k} & 1-a_0 ^{k} \\ 0&0&1\end{pmatrix} \\[6pt]
& A_{DD}^k =\begin{pmatrix}d^{2k}& d(1-d)\frac{a_0^k - d^{2k}}{a_0-d^2} & 1-d^{2k}-d(1-d)\frac{a_0^k-d^{2k}}{a_0-d^2} \\ 0 & a_0^k & 1-a_0^k \\ 0&0&1\end{pmatrix} . \label{eq:ADDk}
\end{align}

With these preliminary calculations, we can determine the $n$-step payoff of $S_j$ versus $S_i$, which will be the focus of our analysis.  We call this payoff $A(S_j;S_i)$ and note that it is equal to the probability the individual with strategy $S_j$ is still alive after the $n$ steps of the game.  For the case $j \geq i$, we have 
\begin{align}
  A(S_j;S_i) =&~ (A_{CC}^{n-j} A_{DC}^{j-i} A_{DD}^i)_{(1,1)}+(A_{CC}^{n-j} A_{DC}^{j-i} A_{DD}^i)_{(1,2)} \notag \\[6pt]
    =&~ a^{2(n-j)}(bc)^{j-i}d^{2i} + a(1-a)a_0^j \frac{a_0^{n-j}-a^{2(n-j)}}{a_0-a^2} \notag \\ 
     &~ + c(1-b)a^{2(n-j)}a_0^{i}\frac{a_0^{j-i} - (bc)^{j-i}}{a_0-bc}+d(1-d)a^{2(n-j)}(bc)^{j-i} \frac{a_0^i-d^{2i}}{a_0 -d^2} . \label{eq:base}
\end{align}
For the case where $j \leq i$, we get the symmetric result in $b$ and $c$, as well as in $i$ and $j$,
\begin{align}
  A(S_j;S_i) =&~ (A_{CC}^{n-i} A_{CD}^{i-j} A_{DD}^j)_{(1,1)}+(A_{CC}^{n-i} A_{CD}^{i-j} A_{DD}^j)_{(1,2)} \notag \\[6pt]
    =&~ a^{2(n-i)}(bc)^{i-j}d^{2j} + a(1-a)a_0^i \frac{a_0^{n-i}-a^{2(n-i)}}{a_0-a^2} \notag \\ 
    &~ + b(1-c)a^{2(n-i)}a_0^{j}\frac{a_0^{i-j}-(bc)^{i-j}}{a_0-bc}+d(1-d)a^{2(n-i)}(bc)^{i-j} \frac{a_0^j-d^{2j}}{a_0 -d^2} . \label{eq:base2} 
\end{align}
Note that each of the four terms in (\ref{eq:base}) and (\ref{eq:base2}) correspond to a particular type of sub-event: the first is when the partner also stays alive during the whole game, the remaining three are when the partner dies either when both players have strategy $C$, when one has $C$ and one has $D$, or when both have $D$.

\section{Playing with a fully cooperative partner} \label{sec:fullcoop}

We begin with the example of an individual with strategy $S_j$ and a partner with strategy $S_0$, first in general then focusing on the Prisoner's Dilemma.  We are motivated by the fact that when the single-step game is a Prisoner's Dilemma, playing $D$ in the final step of an $n$-step game will always increase the survival probability of an individual.  If payoffs accrued additively as in the classical repeated Prisoner's Dilemma \citep{RapoportAndChammah1965,Axelrod1984} then by backward induction the same logic would apply to every preceding step of the game.  Seeing an uninterrupted sequence of increased chances of survival, an all-$C$ individual facing an all-$C$ partner would switch to all-$D$.  But payoffs do not accrue additively in an iterated survival game.  We may infer from the results of \citet{WakeleyAndNowak2019} that increasing numbers of defections may eventually be disfavored even against a fully cooperative partner, in particular if the partner were to die and the cost of having to survive the rest of the game alone was too great.   

Here and throughout, we would like to know what strategy an individual might adopt to maximize survival given the partner's strategy and the specific game parameters $(a,b,c,d,a_0,n)$.  In Section~\ref{sec:fourgames}, we illustrate differences among the four well-known classes of games and highlight the importance of the loner survival probability $a_0$ in determining broad patterns of strategy choice in iterated survival games.  Our focused analysis in Section~\ref{sec:PD vs coop} addresses the question just raised, about how far a notion like backward induction might carry over to iterated survival games in which the single-step game between two players is a Prisoner's Dilemma.  Section~\ref{sec:PD vs coop} also introduces the analytical approaches we will apply to the more complicated case of $S_j$ against $S_i$ in Section~\ref{sec:local} and Section~\ref{sec:global}. 

The $n$-step payoff, or probability of survival, of $S_j$ against $S_0$ is obtained by putting $i=0$ in (\ref{eq:base}): 
\begin{equation}\label{eq:vscoop}
    A(S_j;S_0) = \frac{a-a^2}{a_0-a^2}a_0^n+a^{2n}\left[ \frac{a_0 -c}{a_0 -bc} \left(\frac{bc}{a^2}\right)^j + \left(\frac{c-bc}{a_0 - bc}-\frac{a-a^2}{a_0 -a^2}\right) \left(\frac{a_0}{a^2}\right)^j \right] .
\end{equation}
Thus, $A(S_j;S_0)$ depends on three individual survival probabilities $(a,b,c)$, as well as on the pair survival probabilities $(a^2,bc)$ and the loner survival probability $(a_0)$ which are eigenvalues of the single-step matrices in (\ref{eq:acc}) and~(\ref{eq:adc}).  It does not depend on $d$ because there are no steps in which both players use strategy $D$.  The dependence on $n$ is simple: $A(S_j;S_0)$ tends to zero as $n$ tends to infinity.  Surviving longer is always less likely.  Conveniently for our purposes, $A(S_j;S_0)$ depends on $j$ only through the terms in the brackets, which do not include $n$.  We focus on these terms and treat $n$ implicitly, noting of course that $j \leq n$.  Because the terms in brackets may increase as $j$ increases, it should be noted that $A(S_j;S_0)$ is a probability---it can never exceed 1---and that if $j=n$ and $n$ tends to infinity, $A(S_j;S_0)$ tends to zero.

We wish to know the value of $j \in [0,n]$ which maximizes the survival probability of the individual for a given parameters $(a,b,c,a_0)$.  Although $j$ is discrete, in order to find an optima we treat (\ref{eq:vscoop}) as a continuous function of $j$.  Three cases can occur, because there is at most one change in sign of the slope.  The maximum can be reached when $j=0$, which would happen for example when $a^2>a_0>c$.  Then the fully cooperative behavior has the greatest chance of survival, no matter how many rounds are being played.  Alternatively, the supremum of the function may be in the limit $j\to\infty$.  Then, for large enough $n$, the best $j$ would be $n$.  In this case $S_n$, or all-$D$, would have the greatest chance of survival against $S_0$.  A third case is that the function has a maximum at some intermediate value, specifically at 
\begin{equation}
    j_{opt} = \frac{\ln{\left[ \left( \frac{a(1-a)(a_0-bc)}{(a_0-c)(a_0-a^2)}-\frac{c(1-b)}{a_0-c} \right) \frac{ \ln{ \left(\frac{a_0}{a^2}\right)}}{\ln{\left(\frac{bc}{a^2} \right)}} \right] }}{\ln{\left(\frac{bc}{a_0}\right)}} \label{eq:realjopt}
\end{equation}
which exists when the argument of the logarithm in the numerator is positive.  In this case, there could be an intermediate step in the game which gives the greatest benefit of switching from $C$ to $D$.  The integer-valued optimum $j$ would be one of the integers on either side of the real-valued $j_{opt}$, 
\begin{equation}
    J_{opt} = \left\{ \begin{array}{c}\left\lfloor j_{opt} \right\rfloor\\ \mathrm{or} \\ \left \lceil j_{opt} \right\rceil 
    \end{array} \right.\label{eq:jopt}
\end{equation}
provided that $n > j_{opt}$.  If $n \leq j_{opt}$, then $S_n$ would again be the best strategy against $S_0$. It is remarkable that the optimal $j$ does not depend on $n$ in this third case, as long as $n$ remains larger than $j_{opt}$.  

\subsection{Comparison of the four types of games} \label{sec:fourgames}

\begin{figure}[h]
\centering
\includegraphics[scale=1.0]{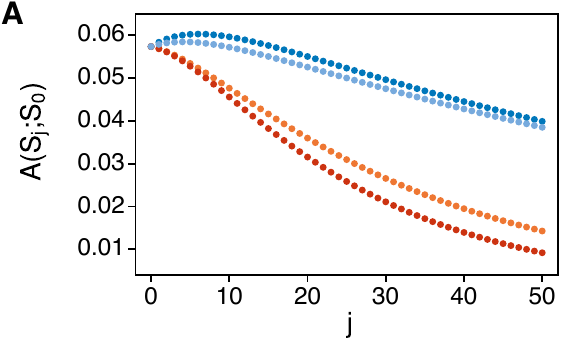} \quad
\includegraphics[scale=1.0]{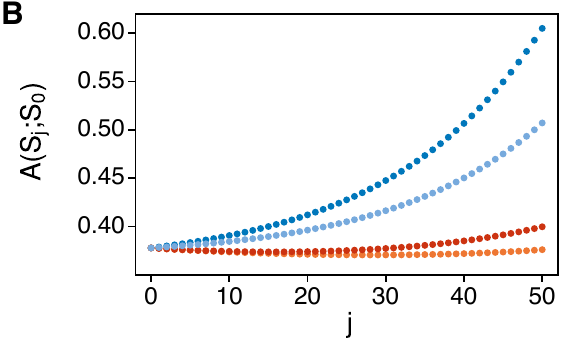}
\caption{The probability of survival of an individual who switches strategy from $C$ to $D$ for the last $j$ steps in of an iterated survival game of length $n=50$ against an all-$C$ partner.  In A the loner survival probability is small, $a_0=0.8$, and in B the loner survival probability is large, $a_0=0.99$.  Colors denote examples of the four possible kinds of games: blue is a Prisoner's Dilemma ($a=0.97$, $b=0.94$, $c=0.99$, $d=0.95$), orange is a Harmony game ($a=0.97$, $b=0.95$, $c=0.96$, $d=0.94$), red is a Stag Hunt ($a=0.97$, $b=0.94$, $c=0.96$, $d=0.95$), light blue is a Hawk Dove game ($a=0.97$, $b=0.95$, $c=0.98$, $d=0.94$).}
\label{fig:SjvsS0}
\end{figure}

Figure~\ref{fig:SjvsS0} shows $A(S_j;S_0)$ as a function of $j$ in a game of length $n=50$ for examples of the four classes of games, when the loner survival probability is either small (Fig.~\ref{fig:SjvsS0}A) or large (Fig.~\ref{fig:SjvsS0}B).  The other payoffs $(a,b,c,d)$ are the same in both panels.  For the example Prisoner's Dilemma, these payoffs ($a=0.97$, $b=0.94$, $c=0.99$, $d=0.95$) are a linear transformation of the classic payoffs ($R = 3$, $S = 0$, $T = 5$, $P = 1$) of \citet{Axelrod1984}.  For all four games in Fig.~\ref{fig:SjvsS0}A the relationship of the eigenvalues is $a^2 > bc > a_0$.  In Fig.~\ref{fig:SjvsS0}B it is $a_0 > a^2 > bc$.  Again, we are interested in whether the highest survival occurs at one or the other extreme, $j=0$ or $j=n$, or at some intermediate $J_{opt}$.  An optimal intermediate strategy exists in these examples only for the Prisoner's Dilemma and the Hawk Dove game with small $a_0$ (Fig.~\ref{fig:SjvsS0}A).  When $a_0$ is the smallest eigenvalue, there is a high cost to a player being alone for a long stretch.  The optimal strategy balances the increased chance of paying this cost against the increase in survival from switching from $C$ to $D$ in the Prisoner's Dilemma and the Hawk Dove game.  If, as in the Stag Hunt and Harmony game in Fig.~\ref{fig:SjvsS0}A, switching from $C$ to $D$ does not directly increase survival, then $S_0$ (all-$C$) is best.  

On the other hand, when $a_0$ is large, a lone individual may have an advantage.  In Fig.~\ref{fig:SjvsS0}B, $a_0$ is the largest payoff and therefore also the largest eigenvalue ($a_0 > a^2 > bc$).  For all four example games, if $j$ is large enough, the term in brackets in (\ref{eq:vscoop}) will be increasing in $j$.  A less-cooperative strategy is advantageous in this case provided the game is long enough.  However, the Harmony game and the Stag Hunt both have $c<a$, so switching from $C$ to $D$ once or a few times directly decreases individual survival causing minima of survival at an intermediate $j$ for both these games.  It is only for larger values of $j$ that the partner's even lower survival ($b<a$ and $b<c$ for all four example games in Fig.~\ref{fig:SjvsS0}) allows the individual to see the benefits of the high loner payoff.  The Prisoner's Dilemma and the Hawk Dove game do not show this dip in survival for small $j$ because they both have $c>a$.  In addition, note that the advantages of increasing $j$ may depend strongly on the partner's survival probability.  For example, changing $b$ so that $b>a$ in the example Harmony game in Fig.~\ref{fig:SjvsS0}B would make increasing $j$ disadvantageous for the individual.  

Figure~\ref{fig:SjvsS0} reveals some key features and some complexities of strategy choice in iterated survival games.  The four-fold classification of games based on the comparison of $a$ to $c$ and $b$ to $d$, together with the rough criteria of large versus small $a_0$ is not enough to determine the potential advantages of switching strategies from $C$ to $D$ at some point in the game.  The order of the eigenvalues $(a^2,bc,d^2,a_0)$ is crucial.   The example games in Figure~\ref{fig:SjvsS0} all have $a^2 > bc > d^2$, but it could be otherwise.  For some games, we might have $a^2 > d^2 > bc$ and for others $bc > a^2 > d^2$.  The assumption that $C$ is the more cooperative and $D$ the less cooperative strategy, hence $a>d$, guarantees that $a^2 > d^2$.  But in all cases, $a_0$ could be anywhere in the order of eigenvalues.  In what follows, we focus on the classic challenge to cooperation, the Prisoner's Dilemma of \citet{Tucker1950} and \citet{RapoportAndChammah1965}, which is a restricted version of what we have been calling the Prisoner's Dilemma class of games.  Our aim is to determine in detail when a late defection might be optimal or when an early one would be better, depending especially on the magnitude of the loner survival probability, $a_0$. 

\subsection{Defection against a fully cooperative partner in the Prisoner's Dilemma} \label{sec:PD vs coop} 

We base our detailed analysis on the payoff difference
\begin{equation}\label{eq:diff vs coop}
    A(S_j;S_0)-A(S_0;S_0) = a^{2n} \left[ \frac{a_{0} - c}{ a_{0}-bc} \left( \frac{bc}{a^2} \right)^{j} + \left( \frac{c(1-b)}{a_0 -bc} -\frac{a(1-a)}{a_0 -a^2} \right) \left( \frac{a_0}{a^2} \right)^{j} - \frac{a_0-a}{a_0 - a^2} \right] . 
\end{equation}
When this difference is positive, there is incentive for an individual (currently playing all-$C$ against an all-$C$ partner) to switch strategies and defect for the final $j$ rounds of the game.  When it is negative, the individual is better off sticking with strategy $S_0$ or switching from $S_j$ to $S_0$.  The $j$ for which this difference is the largest will be the optimal number of end-game defections given a fully cooperative partner.

As in (\ref{eq:vscoop}), there is a separation of $n$ and $j$.  Preliminarily, we may note that (\ref{eq:diff vs coop}) is bounded above by $+1$ and below by $-1$, and that it approaches zero as $n\to\infty$ for any fixed $j$.  Further, the same two exponential terms are present within the brackets, which will increase, decrease or remain constant as $j$ increases depending on the ratios of eigenvalues, $bc/a^2$ and $a_0/a^2$.  So, again, the slope changes sign at most once.  It is straightforward to compute $A(S_0;S_0)-A(S_0;S_0)=0$ and $A(S_1;S_0)-A(S_0;S_0)=a^{2n}(c-a)>0$.  Then for the Prisoner's Dilemma (i.e.\ with $c>a$), the payoff difference increases with $j$ when $j$ is small.  The question is whether it continues to increase or reaches a peak and starts to decrease as $j$ grows.  

To answer this question, we make use of the classical assumptions of the Prisoner's Dilemma, described for example by \citet[p. 34]{RapoportAndChammah1965}.  Specifically,   
\begin{align} 
    &c > a > d > b \label{eq:hyp1} \\
    &a > (b+c)/2 \implies a^2 > bc . \label{eq:hyp2}
\end{align}
The broader class of games which includes this Prisoner's Dilemma is defined just by $c>a$ and $d>b$.  The assumption $a > d$ in (\ref{eq:hyp1}) guarantees that $a^2 > d^2$, which means that the survival probability of the pair is higher when both players cooperate than when both defect.  The additional criterion $a^2 > bc$ in (\ref{eq:hyp2}) means that pairs survive better when both players cooperate than when just one player cooperates.  This is often true: with $(a,b,c,d)$ sampled uniformly at random, $90$\% of survival games which satisfy (\ref{eq:hyp1}) also have $a^2 > bc$ \citep{WakeleyAndNowak2019}.  Meeting this criterion fixes the ratio $bc/a^2$ in (\ref{eq:diff vs coop}) to be strictly less than one.  The parameter $a_0$ remains free, ranging between $0$ and $1$, and the assumptions so far do not determine the relationship between $bc$ and $d^2$. 

With the ratio $bc/a^2<1$, then if it is also true that $a_0 < a^2$, both exponential terms in (\ref{eq:diff vs coop}) will be decreasing in $j$ and will eventually go to zero.  At some point as $j$ increases, assuming $n$ is large enough, the difference $A(S_j;S_0)-A(S_0;S_0)$ will turn negative and converge to the constant $- a^{2n} (a_0-a)/(a_0 - a^2)$.  Too many defections will ultimately hurt the player because the loner survival probability is small.  Again, defecting just once at the end of the game is always favored because $c>a$.  Therefore an optimal strategy $S_{J_{opt}}$ will exist for some integer $J_{opt}$, given by (\ref{eq:realjopt}) and~(\ref{eq:jopt}).  But if $n$ is not large enough, then $j$ will always be less than this optimum and the best strategy against $S_0$ will be $S_n$.

Instead if $a_0 > a^2$, then the difference $A(S_j;S_0)-A(S_0;S_0)$ will eventually be dominated by the middle term in (\ref{eq:diff vs coop}).  Depending on the sign of this term, $A(S_j;S_0)-A(S_0;S_0)$ will be increasing or decreasing when $j$ is large.  As there is at most one change in sign of the slope and the initial slope is positive, the difference $A(S_j;S_0)-A(S_0;S_0)$ will either increase for all $j$ or it will start decreasing at some point as $j$ grows.  Either the best strategy is complete defection or there exists an optimal intermediate strategy.  The first occurs if and only if $\frac{c(1-b)}{a_0 - bc} -\frac{a(1-a)}{a_0 - a^2} >0$, such that the middle term in (\ref{eq:diff vs coop}) is positive.  This induces a cutoff for $a_0$ as it varies between $a^2$ and $1$.  There is a shift in the behavior of $A(S_j;S_0)-A(S_0;S_0)$ as $j$ increases, from having an intermediate optimum to always increasing, specifically at  
\begin{equation} \label{eq:astar}
    a^{\ast}_0 = \frac{c-a}{c-a + a^2-bc} a^2 + \frac{a^2-bc}{c-a + a^2-bc} a . 
\end{equation}
The cutoff $a^{\ast}_0$ is the largest value of $a_0$ such that full defection might not be favored (i.e.\ there is a finite optimum $j$) against a fully cooperative partner.  Again, if $n \leq j_{opt}$, then full defection would still be the best strategy, even if $a_0<a^{\ast}_0$.  But if $a_0>a^{\ast}_0$, then full defection will always favored, for any $n$.  

The two survival differences which determine the coefficients of $a^2$ and $a$ in (\ref{eq:astar}) can be understood with reference to Fig.~\ref{fig:pairtree} and (\ref{eq:acc}) and~(\ref{eq:adc}).  The first, $c-a>0$, is the classic change in payoff for defecting against a cooperative partner, which here is the difference in the single-step survival probability of the individual regardless of what happens to the partner.  The second, $a^2-bc>0$, expresses as a positive term the difference in the probability that both the individual and the partner survive.  It is a single-step cost in pair survival but may be either a cost or a benefit to the individual depending on the values of $a_0$ and $n$.  The coefficients in (\ref{eq:astar}) sum to one, so the cutoff $a^{\ast}_0$ is an average falling between $a^2$ and $a$.  Note that, in view of the first rows of $A_{CC}$ and $A_{DC}$ in (\ref{eq:acc}) and~(\ref{eq:adc}), we may rewrite the shared denominator of these coefficients as the difference in the single-step probability of ending up alone, $c(1-b) - a(1-a) > 0$.  Switching from $C$ to $D$ against an all-$C$ partner increases the chance of winding up alone in every subsequent step of the game, which again may be either a cost or a benefit to the individual.   

The cutoff $a^{\ast}_0$ is closer to $a^2$, and therefore smaller, when the benefit in individual survival, $c-a$, is large relative to the cost in pair survival, $a^2-bc$.  When this is true, even a fairly small value of the loner survival probability $a_0$ cannot prevent $S_n$ from being the best strategy against $S_0$.  On the other hand, $a^{\ast}_0$ is closer to $a$, and therefore larger, when the cost in pair survival is relatively big.  When this is true, there may be an intermediate optimum strategy even when the loner survival probability is fairly large.  Taking derivatives of $a^{\ast}_0$ provides some intuition about the effects of changing specific parameters, when other parameters are held constant.  As long as the assumptions in (\ref{eq:hyp1}) continue to be met, $a^{\ast}_0$ increases as $a$ increases, but decreases when either $b$ or $c$ increases.  In addition, if $b$ increases and $c$ decreases, together so that $bc$ approaches $a^2$, then $a^{\ast}_0$ will decrease toward $a^2$. 

So far, we have considered two possibilities: $a_0 < a^2$ and $a_0 > a^2$.  In the first case, $a^2$ is the largest eigenvalue.  Here a pair of cooperators survives a single step of the game better than any other pair and better than a lone individual.  Both terms which depend on $j$ in (\ref{eq:diff vs coop}) decrease to zero and the payoff difference $A(S_j;S_0)-A(S_0;S_0)$ converges to a finite, negative constant, so there exists an optimum number of end-game defections, $J_{opt}$ in (\ref{eq:jopt}).  In the second case ($a_0 > a^2$), a lone individual survives a single step better than any pair of individuals.  But even when this is true, it is not always advantageous to increase the number of end-game defections.  It is only when $a_0$ exceeds $a^{\ast}_0$, which is larger than $a^2$, that defecting more and more is always favored.  If $a^2 < a_0 < a^{\ast}_0$, there is a $J_{opt}$ which may be relevant depending on the total number of steps in the game, $n$.  Note that when $a_0 = a^{\ast}_0$ there is still a growing interest in defecting, but the dependence on $j$ is different because the middle term in (\ref{eq:diff vs coop}) is equal to zero and $A(S_j;S_0)-A(S_0;S_0)$ converges to a positive constant, $a^{2n} (c-a)/(a^2-bc)$, as $j$ increases. 

\begin{figure}[h]
\centering
\includegraphics[scale=1.0]{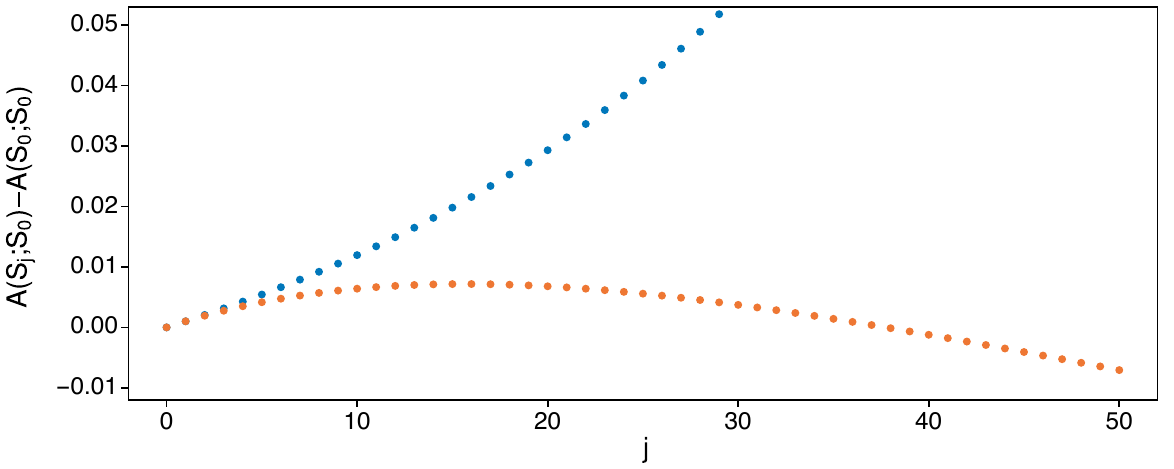}
\caption{$A(S_j;S_0)-A(S_0;S_0)$ as a function of $j$ for two different Prisoner's Dilemmas, illustrating the dependence on $a_0$.  For both: $a=0.97$, $b=0.94$, $c=0.99$, $d=0.95$ as in Fig.~\ref{fig:SjvsS0}, for which $a^2=0.9409$ and $a^{\ast}_{0} \simeq 0.9508$.  In blue: $a_0 = 0.98 > a^{\ast}_0$ and the difference diverges.  In orange: $a_0 = 0.9 < a^2$ and the difference starts to decrease after $J_{opt}=16$.}
\label{fig:optimalvscoop}
\end{figure}

In the special case that $a^2=a_0$, we cannot use the results for geometric series which gave (\ref{eq:ACCk}) through~(\ref{eq:ADDk}).  Here we have
\begin{align}
A(S_j;S_0)&=a^{2n}\left[\frac{a^2-c}{a^2-bc}\left(\frac{bc}{a^2}\right)^{j}+(n-j)\frac{1-a}{a}+\frac{c-bc}{a^2 -bc} \right] \\[6pt]
\frac{\mathrm{d}}{\mathrm{d} j} A(S_j;S_0)&=a^{2n}\left[\frac{a^2-c}{a^2-bc}\ln{\left(\frac{bc}{a^2}\right)}\left(\frac{bc}{a^2}\right)^{j}-\frac{1-a}{a}\right] . 
\end{align}
As $bc<a^2<c$, the derivative will ultimately become negative, so there will be some optimal point of defection.  Thus, $a_0=a^2$ is not pathological and belongs to the case $a^2 < a_0 < a^{\ast}_0$.  For technical reasons we have distinguished three cases --- $a_0 < a^2$, $a^2 \leq a_0 < a^{\ast}_0$ and $a^{\ast}_0 \leq a_0$ --- but the important point is whether a $J_{opt}$ may exist or not, and for this we have just two cases: $a_0 < a^{\ast}_0$ and $a^{\ast} \leq a_0$. Figure~\ref{fig:optimalvscoop} shows the payoff difference, (\ref{eq:diff vs coop}) as function of $j$, for examples of these two cases.

\begin{figure}[b]
\centering
\includegraphics[scale=1.0]{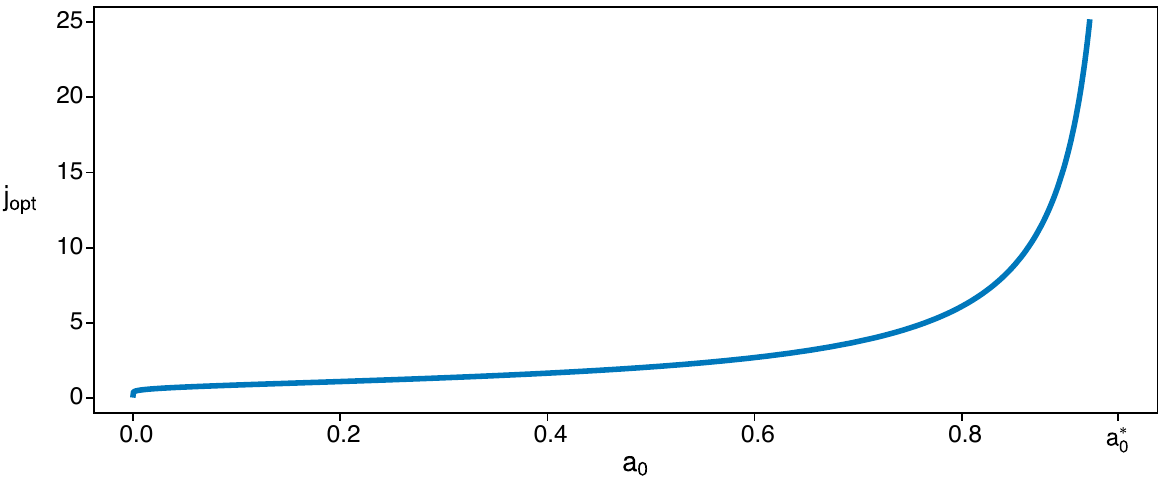}
\caption{The optimal, real-valued point of defection $j_{opt}$ increases without bound as $a_0$ approaches $a^{\ast}_0$, for the same single-step Prisoner's Dilemma game in Fig.~\ref{fig:optimalvscoop}, i.e.\ with $a=0.97$, $b=0.94$, $c=0.99$, $d=0.95$.}
\label{fig:iopt}
\end{figure}

We turn now to the question of how $j_{opt}$ and $J_{opt}$ depend on $a_0$ when $a_0 < a^{\ast}_0$.  Because larger $a_0$ indicates a smaller cost of being alone, it is intuitive that both quantities should increase with $a_0$.  Examination of $j_{opt}$ in (\ref{eq:realjopt}) when $a_0$ is close to either of its extremes, $0$ or $a^{\ast}_0$, gives 
\begin{equation}
    j_{opt} \underset{a_0 \rightarrow{} 0 }{\sim} \frac{\ln{\left( \frac{1-a}{a} \frac{\ln{\left(\frac{1}{a_0}\right)}}{\ln{\left(\frac{bc}{a^2} \right)}}\right)}}{\ln{\left( \frac{1}{a_0} \right)} } \underset{a_0 \rightarrow{} 0}{\rightarrow{}} 0
\end{equation}
and
\begin{equation}
    j_{opt} \underset{a_0 \rightarrow{} a^{\ast}_{0} }{\sim} \frac{\ln{ \left(\frac{ a^{\ast}_{0} -a_0}{(a^{\ast}_0 -c)(a^{\ast}_0 - a^2)}  \frac{\ln{ \left( \frac{a^{\ast}_0}{a^2} \right)}}{\ln{\left( \frac{bc}{a^2} \right)}} \right)} }{\ln{\left( \frac{bc}{a^{\ast}_0 } \right)}} \underset{a_0 \rightarrow{}{a^{\ast}_0} }{\rightarrow} + \infty .
\end{equation}
Figure~\ref{fig:iopt} shows $j_{opt}$ as a function of $a_0$, suggesting that both $j_{opt}$ and $J_{opt}$ are increasing functions of $a_0$. 

For $J_{opt}$, using (\ref{eq:jopt}) and the fact that $A(S_1;S_0)-A(S_{0};S_{0}) = a^{2n}(c-a)>0$, we have 
\begin{equation}
    J_{opt} \underset{a_0 \rightarrow 0} {\rightarrow } 1
\end{equation}
and
\begin{equation}
    J_{opt} \underset{a_0 \rightarrow a^{\ast}_0} {\rightarrow} + \infty .
\end{equation}
To prove that $J_{opt}$ increases with $a_0$ for $a_0 < a^{\ast}_0$, we note that 
\begin{equation}
    J_{opt} = 1 + max \left\{j/ A(S_{j+1};S_0) \geq A(S_j;S_0) \right\} .
\end{equation}
Therefore, it is enough to prove that, for any given $j$, if $A(S_{j+1};S_0) \geq A(S_j;S_0)$ for some $a_0$ then it is also true for any larger $a_0$.  In the special case $j=0$, we have $A(S_1;S_0) \geq A(S_0;S_0)$ for all $a_0$ because in this case (\ref{eq:diff vs coop}) does not depend on $a_0$. Let $j$ be some integer such that $n>j \geq 1$.  Then we have 
\begin{align}
        (A(S_{j+1};S_0)- A(S_j;S_0))(a_0) &= a^{2(n-j-1)} \left(c-a + a^2 -bc \right) a^{j}_0 - a^{2(n-j-1)} (a^2 -bc) (bc)^{j} \notag \\
        &\quad - a^{2(n-j-1)}(a^2 -bc)c(1-b) \sum_{k=0}^{j-1} (bc)^k a^{j-k-1}_{0} \\
        \frac{\mathrm{d~~}}{\mathrm{d} a_0}(A(S_{j+1};S_0)- A(S_j;S_0))(a_0)&= ja^{2(n-j-1)} \left(c-a + a^2 -bc \right) a^{j-1}_0 \notag \\
        &\quad - a^{2(n-j-1)}(a^2 -bc)c(1-b) \sum_{k=0}^{j-2} (j-k-1)(bc)^k a^{j-k-2}_{0} \\
        &\geq \frac{j}{a_0} \Bigg[a^{2(n-j-1)} \left(c-a + a^2 -bc \right) a^{j}_0 \notag \\
        &\quad - a^{2(n-j-1)}(a^2 -bc)c(1-b) \sum_{k=0}^{j-2} (bc)^k a^{j-k-1}_{0} \Bigg]\\
        &\geq \frac{j}{a_0} (A(S_{j+1};S_0)- A(S_j;S_0))(a_0) + \frac{j}{a_0}a^{2(n-j-1)} (a^2 -bc) (bc)^{j} .
\end{align}
Then since $a^2 > bc$, the last inequality completes the proof. When $A(S_{j+1};S_0)- A(S_j;S_0)\geq 0$ for some $a_0$ it remains positive for any larger $a_0$. With an all-$C$ partner, $J_{opt}$ increases with $a_0 < a^{\ast}_0$.  Beyond this point, i.e.\ for $a_0 \geq a^{\ast}_0$, we may also say that $J_{opt}$ is infinite because regardless of $n$ it will always be beneficial to increase the number of defections. 

\section{Behavioral equilibria} \label{sec:local}

We now lift the restriction that the partner is fully cooperative, and ask whether there is an incentive to defect more or to cooperate more when the partner has strategy $S_i$.  As the number of possible strategies $\{S_j$ ; $j \in [0,n]\}$ is finite, there will always be an optimal one against $S_i$.  We are interested in identifying stable strategies, such that the individual cannot increase their probability of survival against a partner who has the same strategy.  Strategy $S_i$ is optimal in this sense, and thus a strict Nash equilibrium, when 
\begin{equation} 
\forall j \neq i,~A(S_i;S_i) > A(S_j;S_i) . \label{eq:nash}
\end{equation}
Due to (\ref{eq:base}) and~(\ref{eq:base2}), the cases $j>i$ and $j<i$ must be analyzed separately.  Note that there may be many such equilibrium strategies.  We will also consider whether these equilibria are evolutionarily stable strategies, or ESSs, \citep{MaynardSmithAndPrice1973,Thomas1985} further satisfying
\begin{equation} 
\forall j \neq i,~A(S_i;S_j) > A(S_j;S_j) . \label{eq:ess}
\end{equation}
Equation~(\ref{eq:ess}) is a population concept: even if an alternative strategy $S_j$ reaches a frequency where its self-interaction becomes appreciable, it will not take over the population.  

In this section we focus on local equilibria, meaning that the only options open to the individual are to defect one more time or cooperate one more time.  Strategy $S_i$ is a locally stable if and only if 
\begin{align} 
& A(S_i;S_i) > A(S_{i+1};S_i) \label{eq:plusone} \\ 
& A(S_i;S_i) > A(S_{i-1};S_i) \label{eq:minusone}
\end{align}
which may be summarized as $A(S_i;S_i)>max(A(S_{i+1};S_i),~A(S_{i-1};S_i))$.  In Section~\ref{sec:global}, we consider global properties of the payoff matrix $A(S_j;S_i)$ for all $i,j \in [0,n]$.

\subsection{General results}

We base our analysis of local stability on the two key differences  
\begin{align}
    A(S_{i+1};S_i)-A(S_i;S_i) &= 
    a^{2(n-1)} \left[(bc-a^2) \frac{a_0 -d}{a_0 -d^2} \left(\frac{d^2}{a^2}\right)^i +\left((a^2 -bc) \frac{a_0 -d}{a_0 -d^2}+c-a\right) \left(\frac{a_0}{a^2}\right)^i\right] \label{eq:i+1} \\[6pt] 
    A(S_{i-1};S_i)-A(S_i;S_i) &= 
    a^{2(n-1)} \left[(bc-d^2) \frac{a_0 -d}{a_0 -d^2} \left(\frac{d^2}{a^2}\right)^{i-1} +\left((d^2 -bc) \frac{a_0 -d}{a_0 -d^2}+b-d\right)\left(\frac{a_0}{a^2}\right)^{i-1}\right] . \label{eq:i-1}
\end{align}
Similar to (\ref{eq:diff vs coop}), these two formulas show a separation of $i$ and $n$. Their signs may depend on $i$ but will not depend on $n$. Both formulas are sums of two exponential functions in $i$, with coefficients that depend on the game parameters $(a,b,c,d,a_0)$.  They can change sign at most once.  Therefore, the conditions for local stability in (\ref{eq:plusone}) and~(\ref{eq:minusone}) will each be met---corresponding, respectively, to (\ref{eq:i+1}) and~(\ref{eq:i-1}) being negative---either for a stretch of $i$ or for no values of $i$. The set of locally stable $i$ is the intersection of these two (possibly empty) stretches. In the case of defecting one more time, the stretch may range from $0$ to $+\infty$. In the case of cooperating one more time, it may range from $1$ to $+\infty$.  Then the locally stable strategies are a stretch of integers whose boundaries range from $1$ to $+\infty$ (which again may be empty) plus possibly $0$.  For the smallest $i$, (\ref{eq:i+1}) and~(\ref{eq:i-1}) reduce to
\begin{align}
   &A(S_1;S_0)-A(S_0;S_0)=a^{2(n-1)}(c-a)\\
   &A(S_0;S_1)-A(S_1;S_1)=a^{2(n-1)}(b-d) .
\end{align}
Strategy $S_0$, or all-$C$, is locally stable if and only if $c<a$ which means that the single-step game is either a Harmony game or a Stag Hunt (cf.\ Table~\ref{tab:twobytwo}). As in Section~\ref{sec:fullcoop}, we treat $n$ implicitly in what follows, keeping in mind that any stretch of equilibria will depend on $n$ in that $n$ fixes the upper boundary of the interval.  Our primary concern is to understand how the stretch of locally stable states depends on the other game parameters, in particular the loner survival probability $a_0$.

\subsection{Focusing on the Prisoner's Dilemma} \label{sec:localPD}

Here as in Section \ref{sec:PD vs coop} we focus on the Prisoner's Dilemma.  Thus we use the exact same assumptions, (\ref{eq:hyp1}) and~(\ref{eq:hyp2}), that $c > a > d > b$ and $a^2 > bc$.  In the following subsections, we first study the incentives (or disincentives) to either defect more or cooperate more, then consider the overlap of these two sets of results in order to identify equilibria, and finally turn to questions about evolutionary stability.   

\subsubsection{Incentives to defect or cooperate more against $S_i$} \label{sec:iDandiC}

Under the assumption that the single-step game is a Prisoner's Dilemma, we have 
\begin{align}
   &A(S_1;S_0)-A(S_0;S_{0})=a^{2(n-1)}(c-a) > 0 \\
   &A(S_0;S_1)-A(S_1;S_1)=a^{2(n-1)}(b-d) < 0.
\end{align}
This proves that $i=0$ is neither a locally stable state nor an ESS when the single-step game is a Prisoner's Dilemma.  The difference $A(S_{i+1};S_i)-A(S_i;S_i)$ in (\ref{eq:i+1}) starts off positive for small $i$ and will change sign at most once.  We define the real-valued cutoff $i_D$ to be the point at which defecting one more time becomes disadvantageous as $i$ increases.  If (\ref{eq:i+1}) never changes sign, then $i_D$ does not exist and additional defection is always favored.   When $i>i_D$, the strategy $S_i$ is a candidate for locally stability.  Similarly, since $A(S_{i-1};S_i)-A(S_i;S_i)$ in (\ref{eq:i-1}) starts off negative for small $i$ and changes sign at most once, we define $i_C$ to be the point at which increased cooperation first becomes advantageous.  Here too $i_C$ may not exist.  When $i<i_C$, the second criterion for local stability of strategy $S_i$ is met.  Both criteria are satisfied when $i \in [ \lceil i_D \rceil ,\lfloor i_C \rfloor ]$, but this interval will be empty if $\lceil i_D \rceil > \lfloor i_C \rfloor$.

We begin with the case of increasing defection.  If $a_0<d^2$, then $A(S_{i+1};S_i)-A(S_i;S_i)$ in (\ref{eq:i+1}) will ultimately become negative because the first term inside the brackets will come to dominate as $i$ grows and this term is negative owing to our assumption that $a^2>bc$. If $a_0>d^2$, then (\ref{eq:i+1}) will ultimately become negative if and only if $(a^2 -bc) \frac{a_0 -d}{a_0 -d^2} +c-a <0$.  Analogous to the situation in Section~\ref{sec:PD vs coop} with the cutoffs $a^{\ast}_0$ and $j_{opt}$, here we require 
\begin{equation}
    a_0 < a^{\prime}_0 = \frac{c-a}{c-a + a^2 -bc} d^2 + \frac{a^2 -bc}{c-a + a^2 -bc} d \label{eq:aprime}
\end{equation}
and find an associated cutoff for $i$ 
\begin{equation}
    i_D=\frac{\ln{\left(1+\frac{c-a}{a^2 -bc}\frac{a_0 -d^2}{a_0-d}\right)}}{\ln{\left(\frac{d^2}{a_0}\right)}} \label{eq:iD}
\end{equation}
which exists if $a_0<a^\prime_0$.  There is an advantage to defecting one more more time only when $i<i_D$. For larger $i$ it is disadvantageous. In the special case $a_0=d^2$, we obtain
\begin{equation}
    A(S_{i+1};S_i)-A(S_i;S_i)=a^{2n}\left[c-a+(bc-a^2)\frac{1-d}{d} i\right] \left(\frac{d^2}{a^2}\right)^i
\end{equation}
which starts off positive for $i=0$ then turns negative for some larger $i$.  Thus $a_0=d^2$ is not a pathological case but belongs with $a_0<d^2$ and $d^2<a_0<a^\prime_0$.  For all $a_0<a^\prime_0$, additional defections will eventually be disadvantageous.  Alternatively, if $a_0 \geq a^{\prime}_0$, an individual with strategy $S_i$ has an incentive to defect one more time against a partner with strategy $S_i$, regardless of the value of $i$.  

\begin{figure}[h]
\centering
\includegraphics[scale=1.0]{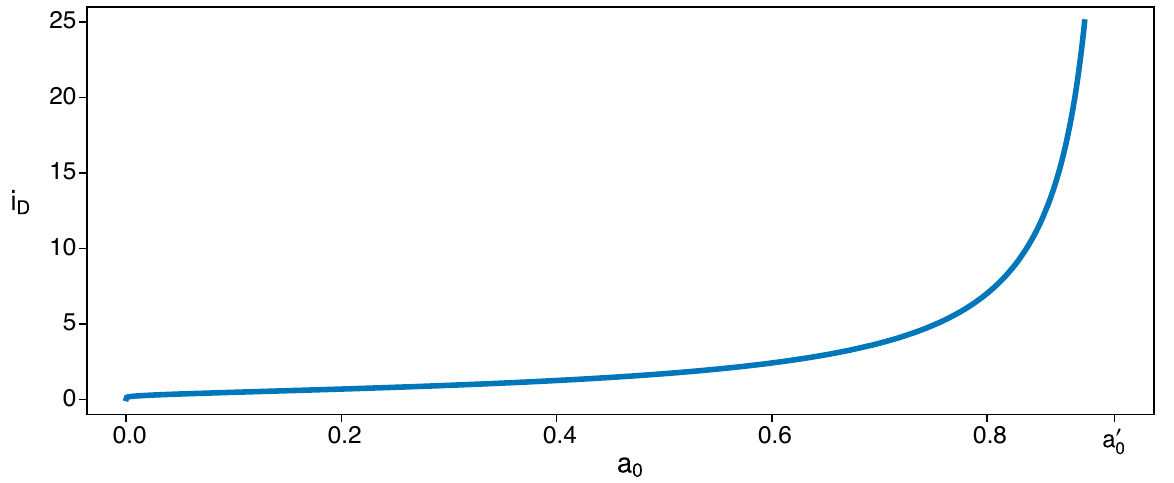}
\caption{$i_D$ is the point above which defecting once more would become disadvantageous.  It increases with $a_0$ toward $+\infty$ as $a_0$ approaches $a^{\prime}_0$. The parameters here are the same as in Fig.~\ref{fig:iopt} ($a=0.97$, $b=0.94$, $c=0.99$, $d=0.95$).}
\label{fig:iD}
\end{figure}

Like $a^\ast_0$ in (\ref{eq:astar}), the cutoff $a^\prime_0$ in (\ref{eq:aprime}) is an average.  Previously $i$ was the number of defections the individual was considering against and all-$C$ partner.  Here $i$ is the fixed number of $DD$ rounds the individual must face when considering whether to defect one more time against an $S_i$ partner.  As a result, $a^\prime_0$ is an average falling between $d^2$ and $d$ instead of between $a^2$ and $a$.  However, the coefficients determining where it falls are the same as before because the individual is making the same switch, from $C$ to $D$ when the partner has strategy $C$ in that step.  By taking derivatives of either coefficient (they sum to one), it can be shown that $a^\prime_0$ is closer to $d$ when $a$ increases, but is closer to $d^2$ when either $b$ or $c$ increases or when $b$ and $c$ together approach $a$.  The effect of $d$ on $a^\prime_0$ is straightforward.  For example, if $d$ is small, then $a^\prime_0$ will be small and both the individual and the partner will have low survival in the remaining steps of the game.  What (\ref{eq:aprime}) and~(\ref{eq:iD}) show is that this can offset the benefit of additional defections.  Although the individual may still see an advantage to increasing defection if $a_0$ is small enough, the advantage will only be realized for $i < i_D$.  Figure~\ref{fig:iD} illustrates that when $a_0$ is small, $i_D$ is small.  

\begin{figure}[h]
\centering
\includegraphics[scale=1.0]{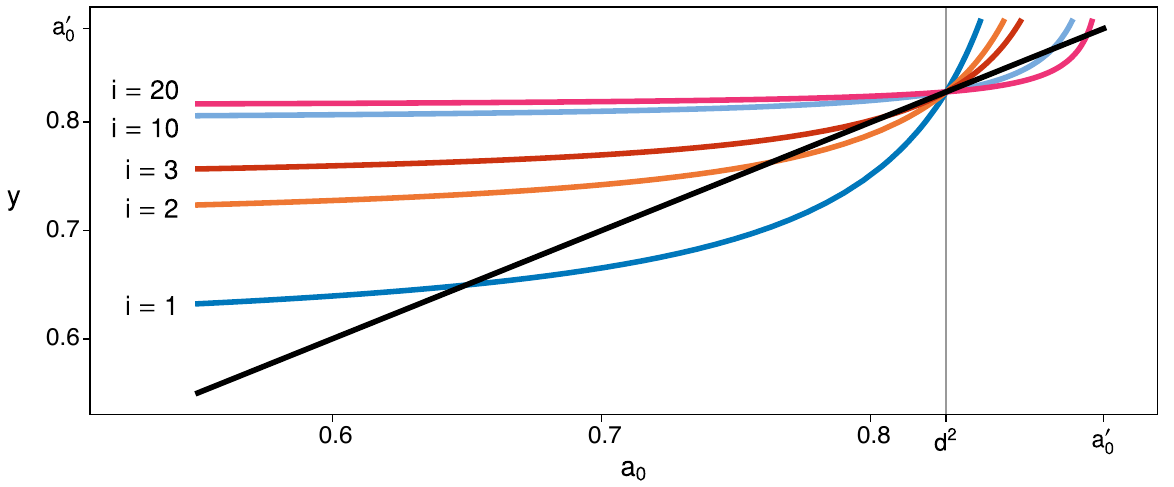}
\caption{Solving (\ref{eq:iDproof}) graphically means finding the intersection between the diagonal $y=a_0$ in black and one of the curves in color $y=d^{2}\left( 1+ \frac{c-a}{a^2 -bc}\frac{a_0 -d^2}{a_0 -d} \right)^{-1/i}$ for a given $i$.  This is illustrated here for five different values of $i$ and game parameters $a=0.97$ , $b=0.9$ , $c=0.99$ and $d=0.91$.  These differ from the parameters in Fig.~\ref{fig:iD} and previous figures by subtracting $0.04$ from $b$ and $d$, which makes $d^2$ smaller while keeping $a_0^\prime$ close to $d$, in order to illustrate the curves in the region $a_0>d^2$.  }
\label{fig:iDproof}
\end{figure}

Figure~\ref{fig:iD} suggests that $i_D$ is an increasing function of $a_0$, growing from $0$ to $+\infty$ as $a_0$ goes from $0$ to $a^\prime_0$.  As before, this fits with intuition about the balance between the benefit of defecting while the partner is still alive and the drawback of having to survive alone. The bigger $a_0$ is, the smaller this drawback becomes.  The extremes of $i_D$ can be obtained from (\ref{eq:iD}).  We find 
\begin{align}
    &i_D \underset{a_0 \rightarrow 0} {\rightarrow } 0 \\
    &i_D \underset{a_0 \rightarrow a^{\prime}_0} {\rightarrow} +\infty .
\end{align}
To prove that $i_D$ is an increasing function of $a_0$, we focus on the point at which defecting one more time switches from being advantageous to being disadvantageous.  This determines the relationship between $i_D$ and $a_0$, namely  
\begin{align}\label{eq:iDproof}
    A(S_{i+1};S_i)-A(S_i;S_i)=0 &\Leftrightarrow (bc-a^2) \frac{a_0 -d}{a_0 -d^2} d^{2i} +\left((a^2 -bc) \frac{a_0 -d}{a_0 -d^2} +c-a\right) a^{i}_0 =0 \notag \\
    &\Leftrightarrow \frac{d^{2}}{\left( 1+ \frac{c-a}{a^2 -bc}\frac{a_0 -d^2}{a_0 -d} \right)^{\frac{1}{i}}}=a_0 .
\end{align}
Both $i = i_D$ and $a_0 = d^2$ are solutions of (\ref{eq:iDproof}).  The solution $a_0 = d^2$ is true for all $i$.  We want to know how the other solution depends on $a_0$, and for this we write $i_D(a_0)$.  We use a graphical method depicted in Fig.~\ref{fig:iDproof}.  Specifically, the two solutions of (\ref{eq:iDproof}) are the two points at which the diagonal $y=a_0$ and the curve $y=d^{2}\left( 1+ \frac{c-a}{a^2 -bc}\frac{a_0 -d^2}{a_0 -d} \right)^{-1/i}$ intersect for a given $i$.  Every one of these curves crosses the diagonal at $a_0=d^2$.  The other point of intersection depends on $i$ and, for each curve, happens at $a_0$ such that $i_D (a_0)$ solves (\ref{eq:iDproof}).  Under the assumptions Eqs~(\ref{eq:hyp1}) and~(\ref{eq:hyp2}), the function $d^{2}\left( 1+ \frac{c-a}{a^2 -bc}\frac{a_0 -d^2}{a_0 -d} \right)^{-1/i}$ increases with $a_0$ and, for a given $a_0<d^2$, it increases with $i$.  Then because these curves are anchored at $a_0=d^2$, the other points at which they cross the diagonal, which we call $a_{0}(i)$, must also increase with $i$.  Considering two values of $i$, with $i_1 > i_2 > 0$, we have 
\begin{equation}
    a_0 < d^2 \Rightarrow \frac{d^{2}}{\left( 1+ \frac{c-a}{a^2 -bc}\frac{a_0 -d^2}{a_0 -d} \right)^{\frac{1}{i_1}}} > \frac{d^{2}}{\left( 1+ \frac{c-a}{a^2 -bc}\frac{a_0 -d^2}{a_0 -d} \right)^{\frac{1}{i_2}}} 
\end{equation}
so that $a_{0}(i_2)<d^2 \Rightarrow a_{0}(i_1) >a_{0} (i_2)~ \mathrm{ and }~ a_{0}(i_2)>d^2 \Rightarrow a_{0}(i_1)>d^2$, and 
\begin{equation}
        a_0 > d^2 \Rightarrow \frac{d^{2}}{\left( 1+ \frac{c-a}{a^2 -bc}\frac{a_0 -d^2}{a_0 -d} \right)^{\frac{1}{i_1}}} <\frac{d^{2}}{\left( 1+ \frac{c-a}{a^2 -bc}\frac{a_0 -d^2}{a_0 -d} \right)^{\frac{1}{i_2}}} 
\end{equation}
so that $a_{0}(i_2)>d^2 \Rightarrow a_{0}(i_1) >a_{0}(i_2)$.  Finally, because $a_0 (i)$ is a positive strictly increasing function, its reciprocal function $i_D (a_0)$ is a strictly increasing function, which is what we set out to prove.

Turning now to the case of increasing cooperation, we recall that $A(S_{i-1};S_i)-A(S_i;S_i)$ in (\ref{eq:i-1}) is negative for the smallest value, $i=1$.   Based just on this consideration, the stretch of possible local equilibria will continue until $A(S_{i-1};S_i)-A(S_i;S_i)$ switches sign and becomes positive at some $i_C$.  If $i_C$ exists, then for any larger $i$ it will be advantageous for the individual to cooperate one more time, specifically in that step of the game when the partner first defects.  Then for all $i>i_C$, strategy $S_i$ cannot be locally stable, whereas for $i<i_C$ it may be locally stable.  We note that if the individual changes strategy from $S_i$ to $S_{i-1}$ against an $S_i$ partner, the pair-survival probability changes from $d^2$ to $bc$, and the individual survival probability changes from $d$ to $b$.  The net effect of the latter is negative ($b-d<0$).  This direct disadvantage to additional cooperation may be offset by increased pair survival, but only if $bc>d^2$.  Again, the assumptions in (\ref{eq:hyp1}) and~(\ref{eq:hyp2}) do not determine the relationship of $bd$ to $d^2$.  It turns out that $60\%$ of Prisoner's Dilemmas defined by (\ref{eq:hyp1}) and~(\ref{eq:hyp2}) have $bc > d^2$  \citep{WakeleyAndNowak2019}.  

When $bc \leq d^2$, the sign of $A(S_{i-1};S_i)-A(S_i;S_i)$ never changes because the net effect on pair survival, $bc-d^2$, is at most zero and will not be able to offset the direct, individual disadvantage of cooperating one more time.  In this case $i_C$ does not exist, so all strategies are candidates for local stability, the upper limit being set only by $n$.  When $bc > d^2$, the sign of the payoff difference may change, giving a finite $i_C$, but this will depend on the loner survival probability.  If $a_0<d^2$, then $A(S_{i-1};S_i)-A(S_i;S_i)$ will eventually become positive.  The case $a_0=d^2$ gives the same result, but is necessary again to compute the difference in probability without using the results for geometric series as we did previously for the condition on $A(S_{i-1};S_i)-A(S_i;S_i)$.  If $a_0>d^2$ the payoff difference will ultimately become positive if and only if $(bc-d^2 ) \frac{a_0 -d}{a_0 -d^2} +d-b >0$.  Overall, additional cooperation is favored when
\begin{equation}
   a_0 < a^{\prime \prime}_0 = \frac{d-b}{d-b + bc - d^2} d^2 + \frac{bc-d^2}{d-b + bc - d^2 } d 
\end{equation}
but only for $i$ greater than 
\begin{equation}
   i_C = 1 + \frac{\ln{\left(1+\frac{d-b}{bc - d^2}\frac{a_0 -d^2}{a_0-d}\right)}}{\ln{\left(\frac{d^2}{a_0}\right)}} . \label{eq:iC}
\end{equation}
Even when the loner survival probability is small, it will still be disadvantageous to cooperate one more time if $i<i_C$. Using an analogous graphical approach to that for $i_D$, it can be shown that $i_C$ is an increasing function of $a_0$ in the interval $(0,a^{\prime \prime}_0)$.  Further, we have \begin{align}
   &i_C \underset{a_0 \rightarrow 0} {\rightarrow } 1 \\
   &i_C \underset{a_0 \rightarrow a^{\prime \prime}_0} {\rightarrow} +\infty . 
\end{align}
Intuitively, the larger $a_0$ is, the lower the danger of a long stretch of mutual defection, so the individual is less inclined to risk a low probability of individual survival ($b$) in a given step for a greater chance of pair survival ($bc$).  As $a_0$ approaches $a^{\prime \prime}_0$, surviving alone no longer becomes a drawback as $i$ increases.  

\subsubsection{Stretches of locally stable strategies} \label{sec:stretches}

The stretch of locally stable strategies is the interval of integers which satisfy the two conditions summarized as $A(S_i;S_i)>max(A(S_{i+1};S_i),~A(S_{i-1};S_i))$.
The interval of integers we are looking for is $[ \lceil i_D \rceil ,\lfloor i_C \rfloor ]$, which is empty when $\lceil i_D \rceil > \lfloor i_C \rfloor$.
There are three different cases to consider.  The first is when $d^2 \geq bc$, such that $i_C$ does not exist regardless of $a_0$.  With an upper limit of $n$, the integer interval begins as $[1,n]$ when $a_0$ is close to $0$, then shrinks to an empty set as $a_0$ increases, because the lower boundary, $\lceil i_D \rceil$, grows without bound as $a_0$ approaches the cutoff $a^{\prime}_0$ in (\ref{eq:aprime}) and does not exist when $a_0 \geq a^{\prime}_0$.  The second and third cases occur under the condition $bc>d^2$, when $i_C$ may exist.  Here, if $a_0$ is close to $0$, then $\lceil i_D \rceil = \lfloor i_C \rfloor =1$, so $S_1$ is the only locally stable strategy for small $a_0$.  When the chance of surviving alone is very small, cooperation will be advantageous except in the final step of the game.  As $a_0$ increases, both  $i_C$ and $i_D$ increase without bound, but with different consequences depending on whether $a^{\prime \prime}_0 < a^{\prime}_0$ or $a^{\prime \prime}_0 > a^{\prime}_0$.

The latter two cases differ owing to the different rates of increase of the two boundaries $\lceil i_D \rceil$ and $\lfloor i_C \rfloor$ as $a_0$ increases.  For simplicity, let us focus on the continuous interval $[i_D ,i_C]$ which has length $i_C - i_D$.  We again treat $n$ implicitly, knowing that the picture will look different depending on whether $n<i_D$, $i_D<n<i_C$ or $n>i_C$.  If $a^{\prime \prime}_0 < a^{\prime}_0$, then $i_C$ diverges before $i_D$ and $i_C - i_D$ will increase as $a_0$ increases. If $a^{\prime \prime}_0 > a^{\prime}_0$, then $i_D$ diverges before $i_C$ and $i_C - i_D$ will decrease as $a_0$ increases.  In this case of shrinking $i_C - i_D$, since $\lceil i_D \rceil = \lfloor i_C \rfloor =1$ when $a_0$ is close to $0$ there will be at most one locally stable state, which will exist over values of $a_0$ for which $[i_D ,i_C]$ contains an integer.  Local stability becomes impossible when $a_0$ is large enough that $\lceil i_D \rceil$ exceeds $\lfloor i_C \rfloor$.

\begin{figure}[h]
\centering
\includegraphics[scale=1.0]{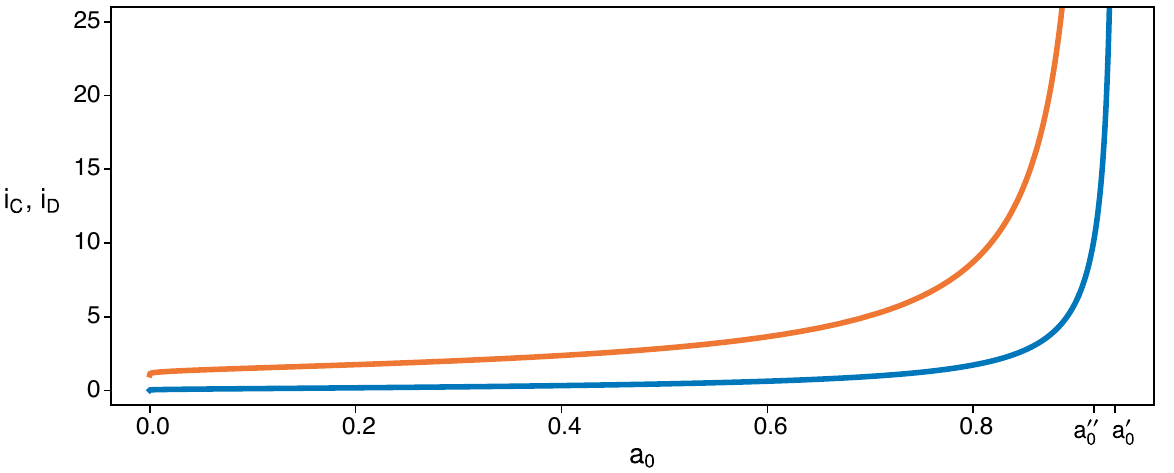}
\caption{In orange, $i_C$ for a given $a_0$ is the point above which an additional round of cooperation is favored.  In blue, $i_D$ for a given $a_0$ is the point below which an additional round of defection is favored.  The game parameters are $a=0.97$, $b=0.93$, $c=0.98$, $d=0.95$, which are related to those used previous, e.g.\ in  in Fig.~\ref{fig:iD}, by subtracting $0.01$ from $b$ and $c$ which makes $a_0^{\prime\prime} < a_0^\prime$ while keeping $bc>d^2$.  For any given $a_0$, the stretch of locally stable states spans vertically between the two lines, where increased cooperation and increased defection are both disfavored.}
\label{fig:enlarging stretch}
\end{figure}

\begin{figure}[h]
\centering
\includegraphics[scale=1.0]{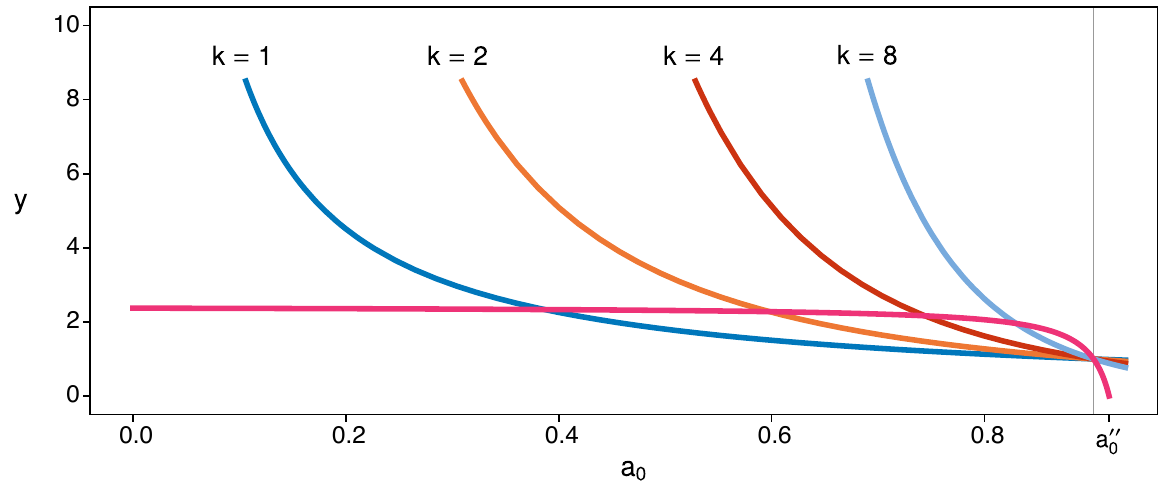}
\caption{Solving (\ref{eq:iCiDproof}) graphically means finding the intersections of the colored curves $y=\left( \frac{d^2}{a_0} \right)^k$ for different values of $k$ and the single black curve $y=\left(1 + \frac{d-b}{bc-d^2} \frac{a_0 -d^2}{a_0 -d}\right)\left(1 +\frac{c-a}{a^2 - bc} \frac{a_0 -d^2}{a_0 -d}\right)^{-1}$.   The parameters are $a=0.97$, $b=0.93$, $c=0.98$, $d=0.95$ as in Fig.~\ref{fig:enlarging stretch}.  All curves intersect when $a_0=d^2$, which is close to $a^{\prime\prime}_0$ in this case and marked by the thin vertical line. }
\label{fig:iCiDproof}
\end{figure}

Figure~\ref{fig:enlarging stretch} illustrates the case where $a^{\prime \prime}_0 < a^{\prime}_0$, so that $i_C$ diverges before $i_D$.  A graphical proof shows that the stretch of equilibria grows with $a_0$ in this case.  Assume some $k>0$.  Then 
\begin{equation}\label{eq:iCiDproof}
        i_C = i_D + k+1 \Leftrightarrow \left( \frac{d^2}{a_0} \right)^k = \frac{1 + \frac{d-b}{bc-d^2} \frac{a_0 -d^2}{a_0 -d}}{1 +\frac{c-a}{a^2 - bc} \frac{a_0 -d^2}{a_0 -d}} .
\end{equation}
As shown in Fig.~\ref{fig:iCiDproof}, graphing the two sides of the right-hand equality in (\ref{eq:iCiDproof}) as functions of $a_0$ shows that the two curves intersect at $a_0=d^2$ regardless of $k$.  This point anchors all the curves, though it is not a permissible solution of (\ref{eq:iCiDproof}) because (\ref{eq:iCiDproof}) was derived assuming $d^2 \neq a_0$.  For any given $k$, the two curves intersect again at another $a_0$ which is the solution of (\ref{eq:iCiDproof}) and which increases with $k$.  We call this value $a_{0}(k)$.  Then for $k_1>k_2 >0$, 
\begin{equation}
   a_0 \leq d^2 \Rightarrow \left( \frac{d^2}{a_0} \right)^{k_1} \geq \left( \frac{d^2}{a_0} \right)^{k_2}
\end{equation}
so we have $a_0 (k_2)\leq d^2 \Rightarrow a_0 (k_1)\geq a_0 (k_2)$ and $a_0 (k_2)> d^2 \Rightarrow a_0 (k_1)>d^2$.  Further, 
\begin{equation}
   a_0> d^2 \Rightarrow  \left( \frac{d^2}{a_0} \right)^{k_1} <\left( \frac{d^2}{a_0} \right)^{k_2} 
\end{equation}
so $a_0 (k_2)> d^2 \Rightarrow a_0 (k_1) >a_0(k_2)$.  Therefore $a_0(k)$ is an increasing function, which means that the bigger the difference between $i_C$ and $i_D$ is, the bigger $a_0$ has to be.  This proves that the length of the equilibrium stretch $i_C - i_D$ increases with $a_0$, approaching infinite length as $a_0$ approaches $a^{\prime \prime}_0$.  When $a^{\prime\prime}_0<a_0<a^{\prime}_0$ the situation is like the first case, $d^2 \geq bc$ which also has infinite $i_C$, and the interval of equilibria $[ \lceil i_D \rceil , n ]$ will shrink until it disappears when $a_0 \geq a^{\prime}_0$.

Thus, with the cap at $n$, the stretch of locally stable equilibria $[ \lceil i_D \rceil ,\lfloor i_C \rfloor ]$ increases in length with its two boundaries drifting towards $n$ as $a_0$ grows.  The upper limit $\lfloor i_C \rfloor$ will reach $n$ for some $a_0 <a^{\prime \prime}_0$ after which the stretch of equilibria will be $[ \lceil i_D \rceil , n ]$ which starts closing as the lower boundary increases with $a_0$. Eventually the stretch will be reduced to the single point $n$ for some $a_0<a^{\prime}_0$. The stretch will disappear as $a_0$ approaches $a^{\prime}_0$, meaning that there will always be an incentive to defect once more.  But since there are only $n$ rounds in the game, $S_n$ will remain a stable strategy for all larger values of $a_0$. 
 
\begin{figure}[h]
\centering
\includegraphics[scale=1.0]{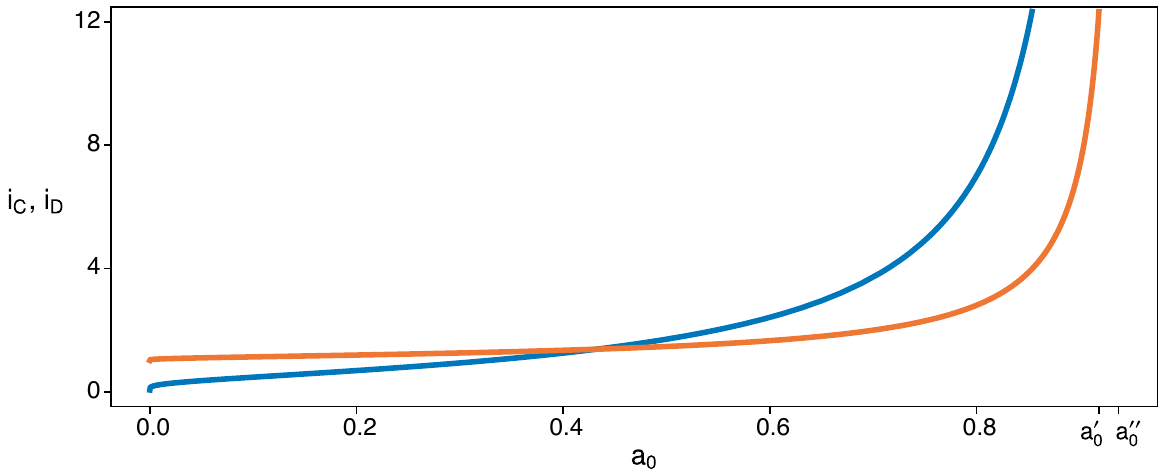}
\caption{As in Fig.~\ref{fig:enlarging stretch}, for a given $a_0$, $i_C$ (orange) is the point above which an additional round of cooperation is favored and $i_D$ (blue) is the point below which an additional round of defection is favored.  The stretch between $i_D$ and $i_C$ shrinks as $a_0$ increases. After the curves cross, the vertical span between $i_C$ and $i_D$ is the interval where both cooperating more and defecting more are better than keeping one's strategy. The curve for $i_D$ is identical to the one plotted in Fig.~\ref{fig:iD} because the same parameters are used here: $a=0.97$, $b=0.94$, $c=0.99$, $d=0.95$.}
\label{fig:shortening stretch}
\end{figure}

Using the same techniques, the opposite behavior can be shown to hold when $a^{\prime}_0 < a^{\prime \prime}_0$.  Specifically, the stretch simply decreases in length, with at most one locally stable state, until it disappears at some $a_0 < a^{\prime}_0$.  Figure~\ref{fig:shortening stretch} shows an example.  For $a_0$ larger than the point where $i_C$ and $i_D$ cross, no stretch of locally stable equilibria can exist.  As long as $n$ is large enough, there will be three zones: for small $i$ there will only be an incentive to defect more, for intermediate $i$ increased defection and increased cooperation will both be favored over keeping the same strategy, and for large $i$ there will only be an incentive to cooperate more.  These three zones will drift towards larger $i$ so that eventually for some $a_0<a^{\prime \prime}_0$ there will only be an advantage to defect one more time.  Then only $S_n$ will remain a stable strategy.

\subsubsection{Summary and interpretation of cases} 

Our analyses in the previous two sections (\ref{sec:iDandiC} and \ref{sec:stretches}) establish that when neither $i_D$ nor $i_C$ exists, there is an incentive to defect one more time against a partner with strategy $S_i$ regardless of $i$.  When $i_D$ exists, additional defections are favored if $i < i_D$ but disfavored if $i > i_D$.  When $i_C$ exists, additional cooperations are disfavored if $i < i_C$ but favored if $i > i_C$.  We focused on the possibility of a non-empty stretch of local equilibria $[\lceil i_D \rceil,\lfloor i_C \rfloor]$ existing when $i>i_D$ and $i < i_C$.  In addition, we described the possibility of a stretch of what we may call `disequilibria', where increased defection and increased cooperation are both favored.  For both kinds of stretches, we established that when $i$ is outside the stretch there is incentive to move toward it by increasing the number of defections if $i<i_D$ and increasing the number of cooperations if $i>i_C$.  Here we point out another possibility, that neither kind of stretch exists, namely when $\lfloor i_D \rfloor = \lfloor i_C \rfloor$ so that increased defection is favored if $i \leq \lfloor i_C \rfloor$ and increased cooperation is favored if $i \geq \lfloor i_C \rfloor + 1$.

Table~\ref{tab:cases} provides further detail and specifies parameter ranges for each case.  Among the ten possibilities listed in Table~\ref{tab:cases}, there are a total of six cases which may be described in terms of the loner survival probability, $a_0$, roughly as follows.  One case holds for large $a_0$, such that additional defections are favored regardless of $i$.  Two cases hold for small $a_0$, such that a stretch of local equilibria is possible which may either be capped by $n$ regardless of how large $n$ is or may be capped by $\lfloor i_C \rfloor$.  However, stretches of equilibria are irrelevant if the game is too short ($n < \lceil i_D \rceil$).  Two more cases hold for some intermediate $a_0$, such that a stretch of local disequilibria is possible which may be capped by $n$ or by $\lfloor i_D \rfloor$, but is irrelevant if the game is too short ($n < \lceil i_C \rceil$).  These intermediate values of $a_0$ occur when $a_0$ is larger than the value for which $i_D=i_C$, which is possible only when $bc>d^2$ and $a_0^{\prime} < a_0^{\prime\prime}$.  We might call this value $a_0^{\prime\prime\prime}$ and for reference give its formula,   
\begin{equation}
   a_0^{\prime\prime\prime} = \frac{b d (c - d) (a^2 - b c)}{(a^2 - b c + c - a) (b c - d^2)} ,
\end{equation}
which may be obtained using (\ref{eq:iD}) and~(\ref{eq:iC}).  For example, $a_0^{\prime\prime\prime} \approx 0.43$ using the parameters of Fig.~\ref{fig:shortening stretch}.  However, the classification of cases for $a_0$ near $a_0^{\prime\prime\prime}$ is complicated because it depends on $\lfloor i_D \rfloor$ and $\lfloor i_C \rfloor$, not simply on $i_D$ and $i_C$.  An additional, sixth case occurs in this region, when $\lfloor i_D \rfloor = \lfloor i_C \rfloor$ such that additional defection is favored if $i \leq \lfloor i_D \rfloor$ and additional cooperation is favored if $i \geq \lceil i_D \rceil = \lceil i_C \rceil$.  Again using the parameters of Fig.~\ref{fig:shortening stretch}, we have $\lfloor i_D \rfloor = \lfloor i_C \rfloor = 1$ for $a_0 \in (0.323,0.547)$. 

\begin{table}[h]
\centering
\begin{tabular}{llcl}
\hline
\multicolumn{3}{l}{$bc \leq d^2$} & \\
& \multicolumn{2}{l}{$a_0 \geq a_0^{\prime}$} & additional defection always favored \\
& \multicolumn{2}{l}{$a_0 < a_0^{\prime}$} & possible stretch of equilibria $[\lceil i_D \rceil,n]$ \\
\multicolumn{3}{l}{$bc > d^2$ and $a_0^{\prime} \geq a_0^{\prime\prime}$} & \\
& \multicolumn{2}{l}{$a_0 \geq a_0^{\prime}$} & additional defection always favored \\
& \multicolumn{2}{l}{$a_0^{\prime\prime} \leq a_0 < a_0^{\prime}$} & possible stretch of equilibria $[\lceil i_D \rceil,n]$  \\
& \multicolumn{2}{l}{$a_0 < a_0^{\prime\prime}$} & possible stretch of equilibria $[\lceil i_D \rceil,\lfloor i_C \rfloor]$ \\
\multicolumn{3}{l}{$bc > d^2$ and $a_0^{\prime} < a_0^{\prime\prime}$} & \\
& \multicolumn{2}{l}{$a_0 \geq a_0^{\prime\prime}$} & additional defection always favored \\
& \multicolumn{2}{l}{$a_0^{\prime} \leq a_0 < a_0^{\prime\prime}$} & possible stretch of disequilibria $[\lceil i_C \rceil,n]$ \\
& \multicolumn{2}{l}{$a_0 < a_0^{\prime}$ and $\lfloor i_D \rfloor > \lfloor i_C \rfloor$} \hspace{4pt} & possible stretch of disequilibria $[\lceil i_C \rceil,\lfloor i_D \rfloor]$ \\
& \multicolumn{2}{l}{$a_0 < a_0^{\prime}$ and $\lfloor i_D \rfloor = \lfloor i_C \rfloor$} & incentives switch between $\lfloor i_D \rfloor$ and $\lceil i_D \rceil$ \\
& \multicolumn{2}{l}{$a_0 < a_0^{\prime}$ and $\lfloor i_D \rfloor < \lfloor i_C \rfloor$} & single equilibrium point $\lceil i_D \rceil=\lfloor i_C \rfloor$ \\
\hline
\end{tabular}
\caption{Parameter regions---determined in large part by the relative magnitude of the loner survival probability $a_0$---which produce different incentives for an individual with strategy $S_i$ to either cooperate once more, defect once more, either or neither, against a partner with the same strategy $S_i$.  It is assumed in all cases that $c > a > d > b$ and $a^2 > bc$.  In the second-to-last line, the incentives switch from favoring additional defection if $i \leq \lfloor i_D \rfloor$ to favoring additional cooperation if $i \geq \lceil i_D \rceil = \lceil i_C \rceil$.}
\label{tab:cases}
\end{table}

Following the discussion of Fig.~\ref{fig:pairtree} in Section~\ref{sec:markov}, we interpret the possibilities outlined in Table~\ref{tab:cases} as a balance between individual survival and pair survival.  The first major division of Table~\ref{tab:cases} has already been discussed. It is based on the assumption that the order of eigenvalues is $a^2 > d^2 \geq bc$, with $a_0$ falling somewhere between 0 and 1.  Here an additional round of cooperation does not benefit the individual ($a-c<0$) or the pair ($bc-d^2 \leq 0$).  Thus the only criterion for stable states is whether additional defections remain favored.  They are favored for small $i$ but become disfavored at some larger value of $i = \lceil i_D \rceil$ which increases with $a_0$.  For $a_0 \geq a^{\prime}_0$ the extent covers all integers and none of the $S_i$ are stable. 

In the second and third major divisions of Table~\ref{tab:cases}, i.e.\ when $a^2 > bc > d^2$, the interval of locally stable states is finite and shifts toward larger integers as $a_0$ increases (cf. Fig.~\ref{fig:enlarging stretch} and Fig.~\ref{fig:shortening stretch}).  As it shifts, its width is either shrinking or extending depending whether $a^{\prime}_0 <a^{\prime \prime}_0$, so that $i_D$ diverges first as in Fig.~\ref{fig:shortening stretch}, or $a^{\prime}_0 > a^{\prime \prime}_0$, so that $i_C$ diverges first as in Fig.~\ref{fig:enlarging stretch}.  Putting this in terms of individual versus pair survival, we have 
\begin{align} \label{eq:condition stretch}
        a^{\prime}_0 < a^{\prime \prime}_0 &\Leftrightarrow \frac{a^2 - bc}{c-a+a^2 -bc}<\frac{bc-d^2}{d-b+bc- d^2} \notag \\
        &\Leftrightarrow (a^2 - bc)(d-b)<(bc - d^2)(c-a) \notag \\
        &\Leftrightarrow \frac{a^2 - bc}{c-a}<\frac{bc-d^2}{d-b} .
\end{align}
Thus, a shrinking stretch of equilibria can occur when the cost to pair survival of an additional defection is small ($a^2-bc \simeq 0$). Then there would not be a big drawback to defecting once more which might outweigh the benefit to individual survival ($c-a$).  Opposition to additional defection would come mainly from the cost of having to survive alone.  Larger $a_0$ would decrease this cost and the lower bound of the stretch of equilibria ($i_D$) would depend strongly on $a_0$.  A shrinking stretch of equilibria can also occur when the cost of additional cooperation is small ($d-b \simeq 0$). Then additional cooperation would not cost much individually and would help the pair ($bc-d^2$), so a big increase in $a_0$ would be needed to make further cooperation unattractive, causing the upper bound ($i_C$) of the stretch of equilibria to grow slowly with $a_0$.  Note that these are the same reasons why there might be a stretch of unstable $S_i$.  For the case $a^{\prime}_0 > a^{\prime \prime}_0$, we would have a similar interpretation of an extending stretch of equilibria, but in terms of $bc \simeq d^2 $ or $c\simeq a$.

\subsubsection{A word about local evolutionarily stable strategies} \label{sec:isolated}

We have shown that long stretches of locally stable strategies are possible.  For example, taking the parameters in Fig.~\ref{fig:enlarging stretch} ($a=0.97$, $b=0.9$, $c=0.99$, $d=0.94$) and setting $a_0=0.9$ gives $[\lceil i_D \rceil,n]=[10,n]$ for a game of any length $n \geq 10$.  But which if any of these might be local ESSs?  Equation~(\ref{eq:ess}) specifies the additional conditions for $S_i$ to be a local ESS, from which we infer 
\begin{align}
    &A(S_i;S_{i+1})> A(S_{i+1};S_{i+1}) \Leftrightarrow i+1>i_C \\
    &A(S_i;S_{i-1})> A(S_{i-1};S_{i-1}) \Leftrightarrow i-1<i_D . 
\end{align}
Therefore
\begin{align}
    i \mathrm{~is~a~local~ESS~} &\Leftrightarrow i\in \left] max(i_D , i_C -1) , min( i_D +1, i_C)\right[ \notag \\
    &\Leftrightarrow i \mathrm{~is~the~only~stable~state.} \label{eq:localess}
\end{align} 
When there is just one locally stable strategy, it is also a local ESS and vice versa.  Note that $n$ may be the only stable state because it is the cap, e.g.\ $n=10$ in the example just given.  Otherwise, single stable states occur when $bc > d^2$ and $a_0$ is not too large (Table~\ref{tab:cases}).  Again, ESS is a population concept.  The implication of (\ref{eq:localess}) is that, even when a long stretch of locally stable strategies exists, a population fixed for a locally stable strategy which is not an ESS is susceptible to invasion by a neighboring strategy.

\section{Global properties of $A(S_j;S_i)$} \label{sec:global}
 
Here we return to the payoff matrix $A(S_j;S_i)$ for all $i,j \in [0,n]$, given by (\ref{eq:base}) for $j \geq i$ and by (\ref{eq:base2}) for $j \leq i$.  To recap: in Section~\ref{sec:fullcoop} we fixed $i=0$ and asked whether an optimal response $j=J_{opt}$ existed, and in Section~\ref{sec:local} we focused on $j=i$ and considered in detail the neighboring states where $j$ and $i$ differ by $1$.  These findings, in particular about $J_{opt}$, $i_D$ and $i_C$, retain their importance in this section, where we study the full payoff matrix $A(S_j;S_i)$.  In the subsections which follow, we investigate the global stability of locally stable strategies, show how $A(S_i;S_i)$ depends on $i$, ascertain key features of a best-response walk on the surface $A(S_j;S_i)$, and extend our findings about evolutionary stability.  As above---again following (\ref{eq:hyp1}) and~(\ref{eq:hyp2})---we continue to assume that the single-step game is a Prisoner's Dilemma.
 
\subsection{Global versus local stability}\label{sec:Global equilibrium}

Global stability is defined as follows:    
\begin{equation}
    S_i \mathrm{~is~a~global~equilibrium} \Leftrightarrow \forall j\neq i ~A(S_i;S_i)>A(S_j;S_i) . 
\end{equation}
This, again, is in the sense of a strict Nash equilibrium.  A globally stable state is obviously a locally stable one.  We will prove that the reciprocal is true.  We consider strategies which either defect $k$ more times or cooperate $k$ more times, compared to a locally stable strategy $S_i$.  From (\ref{eq:base}) and  (\ref{eq:base2}) we have
\begin{align}
A(S_{i+k};S_{i}) =&~ a^{2(n-i-k)} (bc)^{k} d^{2i} + a(1-a) a_0^{i+k} \frac{a_0^{n-i-k}-a^{2(n-i-k)}}{a_0-a^2} \notag \\[6pt]
& + c(1-b) a^{2(n-i-k)} a_0^{i} \frac{a_0^{k}-(bc)^{k}}{a_0-bc} + d(1-d)a^{2(n-i-k)} (bc)^{k} \frac{a_0^i-d^{2i}}{a_0 -d^2} \label{eq:iplusk} \\[8pt]
A(S_{i-k};S_{i}) =&~ a^{2(n-i)} (bc)^{k} d^{2(i-k)} + a(1-a) a_0^{i} \frac{a_0^{n-i}-a^{2(n-i)}}{a_0-a^2} \notag \\[6pt]
& + b(1-c) a^{2(n-i)} a_0^{i-k} \frac{a_0^{k}-(bc)^{k}}{a_0-bc} + d(1-d)a^{2(n-i)} (bc)^{k} \frac{a_0^{i-k}-d^{2(i-k)}}{a_0 -d^2} . \label{eq:iminusk}
\end{align}
From the assumption that the single-step game is a Prisoner's Dilemma, we have $c > a > d > b$ and $a^2 > bc$.  Since we assume $S_i$ is locally stable, we also have  $A(S_{i+1};S_i)<A(S_i;S_i)$ and $A(S_{i-1};S_i)<A(S_i;S_i)$.  

We begin with the case of increasing cooperation.  Specifically, we compare the difference in payoff of two individuals, one who cooperates $k$ additional times and one who cooperates $k-1$ additional times, both having a partner with strategy $S_i$.  Using (\ref{eq:iminusk}) and simplifying, we have 
\begin{align}
A(S_{i-k};S_i) - A(S_{i-k+1};S_i) =&~ a^{2(n-i)} (bc)^{k-1} d^{2(i-k)} \left[ (bc-d^2) \frac{a_0-d}{a_0-d^2} \right. \notag \\[6pt]
& \left. + \left( b-d - (bc-d^2) \frac{a_0-d}{a_0-d^2} \right) \left( \frac{a_0}{d^2} \right)^{i-k} \right] . \label{eq:kkminusone}
\end{align}
Here $k$ ranges from $1$ to $i$. Equation~(\ref{eq:kkminusone}) is negative when $k=1$, due to local stability, and will change sign at most once as $k$ increases from $1$ to $i$.  We need only check the endpoint, $k=i$, where we find 
\begin{equation}
A(S_0;S_i) - A(S_1;S_i) = a^{2(n-i)} (bc)^{i-1} (b-d) < 0 . \label{eq:j0plus}
\end{equation}
Therefore, no additional number of cooperations is favorable against a locally stable strategy. 

In the case of increasing defection, we compare the payoff of an individual who defects $k+1$ times to that of individual who defects $k$ times, against a partner with strategy $S_i$.  Here $k$ ranges from $0$ to $n-1$, but because $n$ may take any value greater than or equal to one we must consider all $k \geq 0$.  Using (\ref{eq:iplusk}) and simplifying, we may write this difference as 
\begin{equation}\label{eq:kkplusone}
A(S_{i+k+1};S_i) - A(S_{i+k};S_i) = a^{2(n-i-k-1)} a_0^i (bc)^{k} \left[ H + (c-a+a^2-bc) \frac{a_0-a_0^\ast}{a_0-bc} \left( \frac{a_0}{bc} \right)^k \right] 
\end{equation}
in which $a_0^\ast$ is the cutoff given by (\ref{eq:astar}), which was derived in the consideration of an optimal number of defections against a partner with strategy $S_0$, and 
\begin{equation}\label{eq:H}
H = (a^2-bc) \left( \frac{c(1-b)}{a_0-bc} - \frac{d(1-d)}{a_0-d^2} - \frac{a_0-d}{a_0-d^2} \left(\frac{d^2}{a_0}\right)^i \right) ,
\end{equation}
which does not depend on $k$.  Local stability means that (\ref{eq:kkplusone}) is negative when $k=0$.  If it remains negative for all $k>0$, then no additional defections will be favored against a partner with strategy $S_i$.  This will depend on the comparison of $H$ and the second term inside the brackets in (\ref{eq:kkplusone}).  If $a_0<bc$, this second term is positive, so from $k=0$ we know $H$ must be negative.  Also, the second term will shrink to zero as $k$ increases because $a_0/bc<1$.  Therefore, the whole of (\ref{eq:kkplusone}) remains negative for all $k$ if $a_0<bc$.  Alternatively, if $bc < a_0 < a_0^\ast$, then the second term in (\ref{eq:kkplusone}) is negative and increases in absolute value as $k$ increases.  Here too (\ref{eq:kkplusone}) remains negative for all $k$.  We do not need to consider $a_0>a_0^\ast$ because local stability requires $a_0 < a_0^\prime$ and we have $a_0^\prime \leq a_0^\ast$.  Thus, we have shown that if $S_i$ is locally stable, there is no increased number of defections which is better.

Taking both cases together, we have proven that locally stable states and globally stable states are the same.  For brevity, we have omitted the detailed treatments of special cases, such as $a_0=a^2$, and simply note that these do not alter our conclusion.  In sum, globally stable states form the same intervals as locally stable states we described previously in Section~\ref{sec:stretches}.  This extension from the local to the global perspective does not necessarily work for an ESS, as we discuss in Section~\ref{sec:ESSglob}.

\subsection{The diagonal $A(S_i;S_i)$} \label{sec:diagonal}

Although potentially long stretches of local equilibria may exist, not all $A(S_i;S_i)$ are equivalent.  In the single-step survival game or in the usual Prisoner's Dilemma with $a>d$, $C$ is a better choice than $D$ if both players take the same strategy.  Here we interested in whether $S_0$ is the best strategy in this sense in the $n$-step game.  We base our analysis on the one-step difference 
\begin{align}
  A(S_{i+1};S_{i+1})-A(S_i;S_i) 
    &= a^{2(n-i-1)} \left[(d^2 -a^2) \frac{a_0 -d}{a_0 -d^2} d^{2i} + \left( (a^2 -d^2) \frac{a_0 -d}{a_0 -d^2}+d-a\right) a_0^i \right] \label{eq:onestep1} \\[6pt]
    &= a^{2(n-i-1)} a_0^i (d-a) \left[ 1 + (a+d) \frac{a_0 -d}{a_0 -d^2} \left( \left(\frac{d^2}{a_0}\right)^i - 1 \right) \right] . \label{eq:onestep2} 
\end{align}
For the smallest $i$ we have 
\begin{equation}
  A(S_1;S_1)-A(S_0;S_0) = a^{2(n-1)} (d-a) < 0 . \label{eq:fad} 
\end{equation}
The difference $A(S_{i+1};S_{i+1})-A(S_i;S_i)$ will remain negative for larger $i$ unless the second term in the brackets in (\ref{eq:onestep2}) becomes too large in the negative direction.  Of course $a+d>0$.  This second term in the brackets is a decreasing function of $a_0$, which begins positive for $0<a_0<d$, then becomes negative when $a_0>d$ and continues to decrease as $a_0$ approaches $1$.  It is straightforward to check that even with $a_0=1$, $A(S_{i+1};S_{i+1})-A(S_i;S_i)$ in (\ref{eq:onestep2}) is negative.  Thus, $A(S_i;S_i)$ is a decreasing function of $i$.  The fully cooperative strategy $S_0$ is the best if both players are restricted to having the same strategy.  

\subsection{A best-response walk on the surface $A(S_j;S_i)$} \label{sec:thewalk}

To better understand the full payoff matrix $A(S_j;S_i)$ for all $i,j \in [0,n]$, we studied the best-response dynamics of an individual who adopts a new strategy which maximizes their survival given their partner's current strategy, and the partner follows suit.  Alternatively, one might think of a larger population, all members of which currently have the same strategy, and in which individuals independently formulate their best response then all switch to that new strategy.  The same procedure is repeated forever.  We will assume that the resulting walk is well defined in the sense that none of the $A(S_j;S_i)$ are equal, considering all $j \in [0,n]$ for a given $i$.  Because the walk is deterministic and has a finite number of possibilities (there are exactly $n+1$ states: $S_0$, $S_1$, $\ldots$, $S_n$), it cannot be injective.  Ultimately the walk will end in a cycle, which might consist of a just one globally stable strategy.

Best-response dynamics show how individuals search for and find pure Nash equilibria when they exist \citep{Roughgarden2016}.  The stretches of stable strategies described in Section~\ref{sec:stretches} and Section~\ref{sec:Global equilibrium} are sets of pure Nash equilibria.  The analysis of $i_D$ and $i_C$ based on single-step changes in strategy (see Section~\ref{sec:localPD}) shows that there is incentive to move toward such a stretch of equilibria for any partners' or prevailing strategies outside the stretch, by increasing defection when $i \leq \lfloor i_D \rfloor$ and by increasing cooperation when $i \geq \lceil i_C \rceil$.  The same analysis shows that there is incentive to move similarly toward a stretch of disequilibria which is not capped by $n$ or a stretch of equilibria which is empty.  Here we investigate how best-response walks on the surface $A(S_j;S_i)$ depend on the initial value of $i$, how stretches of equilibria or disequilibria are approached from above and below in steps which may be greater than one, and how these walks either converge on single points (i.e.\ pure Nash equilibria) or enter into larger cycles.

\begin{figure}[h]
\centering
\includegraphics[scale=1.0]{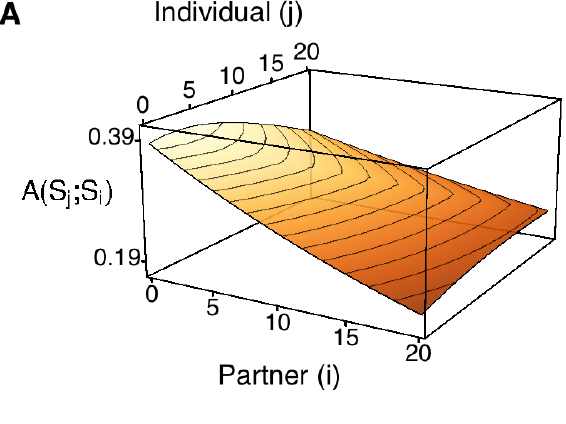} $\quad$ \includegraphics[scale=1.0]{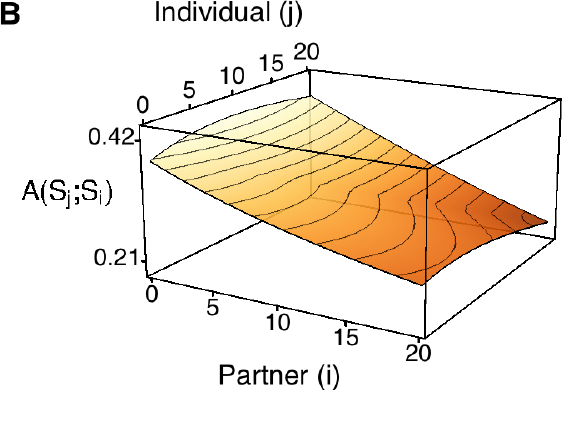} \\ \smallskip
\includegraphics[scale=1.0]{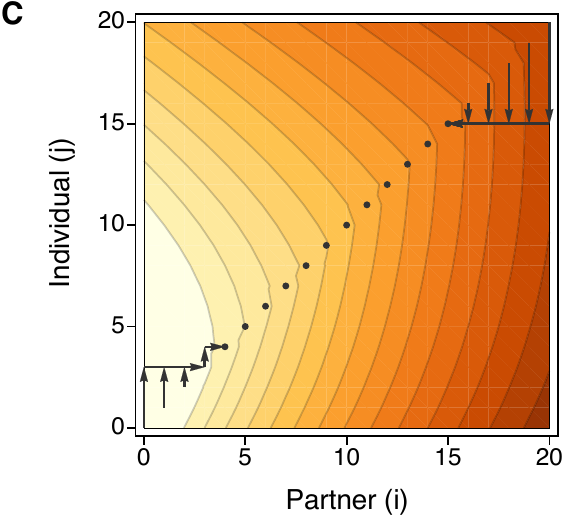} $\quad$ \includegraphics[scale=1.0]{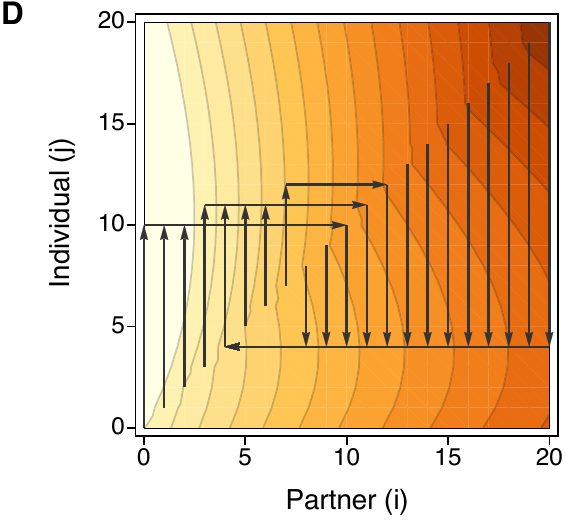} 
\caption{Panels A and B show two payoff surfaces, $A(S_j;S_i)$, for a game of length $n=20$. In both: $a_0=0.86$.  In A: $(a,b,c,d)=(0.97,0.93,0.98,0.95)$ as in Fig.~\ref{fig:enlarging stretch}.  In B: $(a,b,c,d)=(0.97,0.94,0.99,0.95)$ as in Fig.~\ref{fig:shortening stretch}.  Panels C and D show the different possible best-response walks on the same two surfaces.  In C (and A), $i_D = 3.08$ and $i_D = 15.99$ and there is a stretch of equilibria for $i=4$ to $i=15$ which is approached from above and below.  In D (and B) $i_C = 4.22$ and $i_D = 13.49$ and there is a stretch of disequilibria for $i=5$ to $i=13$, leading in this case to a two-state cycle between $i=4$ and $i=11$.} \label{fig:standard walk} 
\end{figure}

Figure~\ref{fig:standard walk} illustrates this for two survival games of length $n=20$, one with a stretch of equilibria and one with a stretch of disequilibria, in which each single step is a Prisoner's Dilemma.  The first (Fig.~\ref{fig:standard walk}AC) has $a_0 < a_0^{\prime\prime} < a_0^{\prime}$ and $0 < \lceil i_D \rceil < \lfloor i_C \rfloor < n$ and so exemplifies the fifth of the ten possibilities listed in Table~\ref{tab:cases}, with a stretch of equilibria for $i \in [4,15]$.  The second (Fig.~\ref{fig:standard walk}BD) has $a_0 < a_0^{\prime} < a_0^{\prime\prime}$ and $0 < \lceil i_C \rceil < \lfloor i_D \rfloor < n$ and so exemplifies the eighth of the ten possibilities listed in Table~\ref{tab:cases}, with a stretch of disequilibria for $i \in [3,13]$.  Panels A and B give 3d depictions of $A(S_j;S_i)$ as a continuous surface.  Panels C and D show the same surfaces, viewed from above, and display all possible best-response walks using arrows.  Each possible walk starts at some point on the diagonal.  It follows the vertical arrow which goes either up or down to the optimal strategy $S_j$ against $S_i$.  Then it follows the horizontal arrow which goes back to the diagonal.  It continues in like manner, repeating the exact same procedures.

Figure~\ref{fig:standard walk} shows the characteristic features of walks when $i_D$ and $i_C$ exist.  In particular, if $i \leq \lfloor i_D \rfloor$ the best response is an increasing function of $i$, whereas if $i \geq \lceil i_C \rceil$ the best response does not depend on $i$.  When there is a stretch of equilibria, $[\lceil i_D \rceil,\lfloor i_C \rfloor]$, the points on the interior are their own best responses, and walks which begin outside the stretch converge on its endpoints, $\lceil i_D \rceil$ from below and $\lfloor i_C \rfloor$ from above.  When there is a stretch of disequilibria, $[\lceil i_C \rceil,\lfloor i_D \rfloor]$, incentives to defect more send walks into the interior then through the stretch, toward $\lceil i_D \rceil$, but these are opposed by incentives to cooperate more, which always leap over the stretch, directly to $\lfloor i_C \rfloor$.  In this case, walks may converge on cycles of two or more states.

We can use (\ref{eq:kkminusone}) and (\ref{eq:kkplusone}) in Section~\ref{sec:Global equilibrium} to obtain the best responses for $i \geq \lceil i_C \rceil$ and $i \leq \lfloor i_D \rfloor$, respectively.  In the first case, we put $j=i-k$ in (\ref{eq:kkminusone}) and rewrite it for our purposes here as  
\begin{align}
A(S_j;S_i) - A(S_{j+1};S_i) =&~ a^{2(n-i)} (bc)^{i-j-1} d^{2j} \left[ (bc-d^2) \frac{a_0-d}{a_0-d^2} \right. \notag \\[6pt]
& \left. - (b - d + bc - d^2) \frac{a_0-a_0^{\prime\prime}}{a_0-d^2} \left( \frac{a_0}{d^2} \right)^{j} \right] . \label{eq:kkminusonebest}
\end{align}
Now $j$ ranges from $0$ to $i$.  We know that (\ref{eq:kkminusonebest}) is negative when $j=0$, from (\ref{eq:j0plus}) which holds for all $i$.  In addition, because here we are assuming $i \geq \lceil i_C \rceil$, we know that (\ref{eq:kkminusonebest}) is positive when $j=i$.  We treat $j$ as continuous and solve for the value which makes (\ref{eq:kkminusonebest}) equal to zero, 
\begin{equation}
    j^{\ast} = \frac{ \ln{ \left( \frac{bc-d^2}{b-d+bc-d^2}\frac{a_0-d}{a_0-a_0^{\prime\prime}} \right) } }{\ln{\left(\frac{a_0}{d^2}\right)}} . \label{eq:jstar}
\end{equation}
Then, the best response falls in the interval $(j^{\ast},j^{\ast}+1)$ and must be equal to $\lceil j^{\ast} \rceil$.  Writing (\ref{eq:jstar}) in this way emphasizes that we are considering the case $a_0<a_0^{\prime\prime}$, namely when $i_C$ exists.  In fact, it is straightforward to show that $j^{\ast} = i_C - 1$, so that $\lceil j^{\ast} \rceil = \lfloor i_C \rfloor$.  Thus, for partner or prevailing strategies with $i \geq \lceil i_C \rceil$, the optimal strategy of an individual is to defect only in the final $\lceil j^{\ast} \rceil = \lfloor i_C \rfloor$ steps of the game.  If there is a stretch of equilibria then $\lceil j^{\ast} \rceil$ is at the upper end of the stretch, whereas if there is a stretch of disequilibria then $\lceil j^{\ast} \rceil$ is just beyond the lower end of the stretch. 

In the second case, $i \leq \lfloor i_D \rfloor$, we similarly set (\ref{eq:kkplusone}) equal to zero and solve to obtain 
\begin{equation}
    k^{\ast}(i) = \frac{ \ln{ \left( \frac{-H(a_0-bc)}{(c-a+a^2-bc) (a_0-a_0^\ast)} \right) } }{\ln{\left(\frac{a_0}{bc}\right)}} \label{eq:kstari}
\end{equation}
in which the dependence on $i$ is through $H$, given by (\ref{eq:H}).  The best response is captured by the interval $(i+k^{\ast}(i),i+k^{\ast}(i)+1)$ and is equal to $i + \lceil k^{\ast}(i) \rceil$.  The full expression for $k^{\ast}(i)$ is cumbersome, but for the smallest $i$ we have 
\begin{equation}
    k^{\ast}(0) = \frac{ \ln{ \left( \frac{a^2-bc}{c-a+a^2-bc}\frac{a_0-c}{a_0-a_0^{\star}} \right) } }{\ln{\left(\frac{a_0}{bc}\right)}} . \label{eq:kstar0}
\end{equation}
Note that this is another route to the optimal number of defections against a fully cooperative partner (Section~\ref{sec:PD vs coop}) because $\lceil k^{\ast}(0) \rceil= J_{opt}$.  For larger $i$, we find that $k^{\ast}(i)$ decreases with $i$, finally reaching zero for $i=i_D$.  As Fig.~\ref{fig:standard walk} shows, the optimal total number ($i + \lceil k^{\ast}(i) \rceil$) of end-game defections against partner or prevailing strategies with $i \leq \lfloor i_D \rfloor$ increases with $i$.  The largest integer-valued $i$ which still favors increased defection is $i = \lfloor i_D \rfloor$ and this would motivate one additional defection by the individual, up to $j = \lceil i_D \rceil$.  If there is a stretch of equilibria, this largest value is at the lower end of the stretch, whereas if there is a stretch of disequilibria it is just beyond the upper end of the stretch.  However, in the latter case, as the walk moves through the stretch, it may happen as in Fig.~\ref{fig:standard walk}D that it never reaches $j = \lceil i_D \rceil$ and instead turns downward because there is an even stronger incentive for additional cooperation.

\begin{figure}[h]
\centering
\includegraphics[scale=1.0]{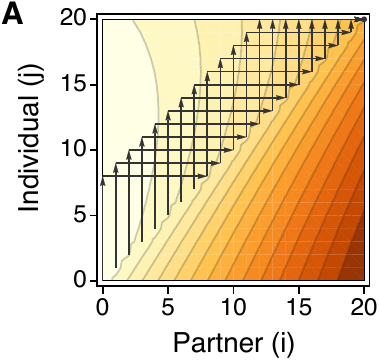} $\quad$ 
\includegraphics[scale=1.0]{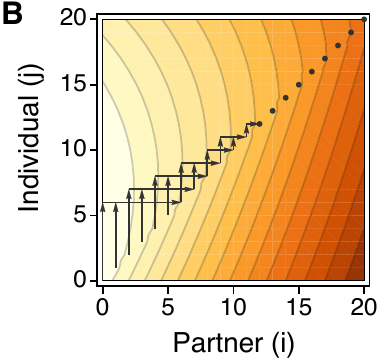} $\quad$ 
\includegraphics[scale=1.0]{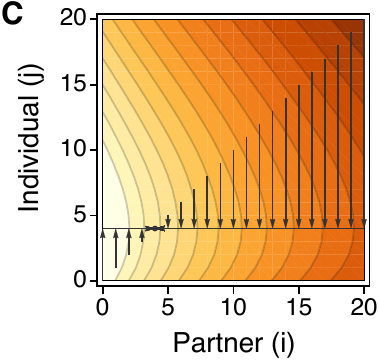} 
\caption{Three additional examples of best-response walks on the surface $A(S_j;S_i)$ for a game of length $n=20$.  In A, $(a,b,c,d)=(0.97,0.91,0.98,0.95)$ and $a_0=0.95$, so $bc<d^2$ and $a_0 > a_0^\prime$, and additional defection is always favored. In B, $(a,b,c,d)=(0.97,0.93,0.98,0.95)$ and $a_0=0.92$, so $bc>d^2$ and $a_0^{\prime\prime} < a_0 < a_0^\prime$, and all $i \geq 12$ are stable.  In C, $(a,b,c,d)=(0.97,0.93,0.99,0.949)$ and $a_0=0.78$, so $bc>d^2$, $a_0 < a_0^{\prime} <  a_0^{\prime\prime}$ and $\lfloor i_D \rfloor < \lfloor i_C \rfloor$, and there is a single stable state at $i = 4$.  Thus, these correspond to the first, third and last of the ten possibilities listed in Table~\ref{tab:cases}. } \label{fig:3morewalks}
\end{figure}

The examples in Fig.~\ref{fig:standard walk} represent just two of the six distinct outcomes among the ten total possibilities listed in Table~\ref{tab:cases}, namely when there is either a stretch of equilibria or a stretch of disequilibria and, in these particular examples, when $n$ is large enough that the entire stretch is apparent within the game.  Figure~\ref{fig:3morewalks} shows three more of the six outcomes: a case in which additional defection is favored for all $i$ (Fig.~\ref{fig:3morewalks}A), a case in which there is a stretch of equilibria capped by $n$ (Fig.~\ref{fig:3morewalks}B), and a case in which there is necessarily a single equilibrium point (Fig.~\ref{fig:3morewalks}C).  These are the first, fourth, and tenth of ten possibilities listed in Table~\ref{tab:cases}. The remaining outcome of the six, which is the ninth possibility in Table~\ref{tab:cases}, when incentives switch between $\lfloor i_D \rfloor = \lfloor i_C \rfloor$ and $\lceil i_D \rceil = \lceil i_C \rceil$, is not depicted but will result in a cycle between those two adjacent states. 

Our findings about $k^{\ast}(i)$ and $j^\ast$ can be applied to all cases, separately for $j \geq i$ above the diagonal and $j \leq i$ below the diagonal.  The optimum $k^{\ast}(i)$ is an extension of $J_{opt}=k^{\ast}(0)$, with the intuitive conclusion that if $i_D$ exists then, as the partner defects more, there is a diminishing return on additional defections by the individual.  Figure~\ref{fig:3morewalks}A shows a case when $i_D$ does not exist and there is no diminishing return on additional defection as $i$ increases. 

In fact, there may still be a diminishing return when $i_D$ does not exist, specifically if $a_0 < d$.  But if $a_0=d$ as in Fig.~\ref{fig:3morewalks}A then $k^{\ast}(i) = J_{opt}$ for all $i$, and if $a_0>d$ then $k^{\ast}(i)$ increases with $i$.  To prove these statements, first it can be shown that 
\begin{equation}
 A(S_{i+k+1},S_i)-A(S_{i+k},S_i) =  \frac{a_0}{a^2} \left[A(S_{i+k+1},S_i)-A(S_{i+k},S_i)\right] +\frac{(a_0 -d)(a^2-bc)}{a^2}(bc)^k d^{2i} . 
\end{equation}
Then, because $k^{\ast}(i)= \max\left\{k>1 | A(S_{i+k},S_i) > A(S_{i+k-1},S_i)\right\} = \min \left\{k>1 | A(S_{i+k+1},S_i) < A(S_{i+k},S_i)\right\}$, we have that $k^{\ast}(i)$ increases with $i$ if $a_0 >d$, decreases if $a_0 <d$ and is constant if $a_0=d$.  As an immediate consequence, we have that $i+ k^{\ast}(i)$ increases with $i$ for $a_0 \geq d$.  We can prove the same is true for $a_0 < d$, in particular for any $i\leq \lceil i_D \rceil$ (which we note might be infinite).  We fix $i\leq \lceil i_D \rceil $ and use $l=i+k$ such that $\lceil i_D \rceil > l >i$. Let $S_j^{(l)}$ be a strategy ending with $j$ defections in a subgame of only $l$ rounds.  We have
\begin{align}
    &A(S_{l+1},S_{i+1})-A(S_l,S_{i+1})-\left[A(S_{l+1},S_i)-A(S_l,S_i)\right]= \notag \\
    &\quad=(bc -a^2)a^{2(n-l-1)} \left(A(S^{(l)}_l,S^{(l)}_{i+1})-A(S^{(l)}_l,S^{(l)}_i) \right) \notag \\
    &\quad=(bc -a^2)a^{2(n-l-1)} \left[\left(A(S^{(l)}_l,S^{(l)}_{l-1}) - A(S^{(l)}_{l-1},S^{(l)}_{l-1}) \right)+\left(A(S^{(l)}_{l-1},S^{(l)}_{l-1}) - A(S^{(l)}_l,S^{(l)}_{l}) \right)\right] . \label{eq:increasingipluskstari}
\end{align}
The second term in the brackets in (\ref{eq:increasingipluskstari}) is always positive thanks to the diagonal behavior described in Section~\ref{sec:diagonal}.  The first term in the brackets is also positive, because $l-1 <i_D$, meaning there is a local incentive to defect one more time.  Note, we used the fact that $i_D$ is does not depend of the number of rounds in the game ($l$ or $n$).  To finish the proof, we further note that $k^{\ast}(i)= \max \left\{k>1 | A(S_{i+k},S_i) < A(S_{i+k-1},S_i)\right\}$.  In sum, optimal defection steps always lead to more total defection.

The result $j^\ast$ may be even more surprising.  It says that when $i_C$ exists, then for any partner strategies with $i \geq \lceil i_C \rceil$---that is, even against a partner who defects in every step of an arbitrarily long game---the optimal strategy is to cooperate for the first $n-\lfloor i_C \rfloor$ steps then defect just $\lfloor i_C \rfloor$ times at the end of the game.  Whereas $J_{opt}$ shows the limitation of backward induction in iterated survival games, $j^\ast$ for the case $i=n$ shows the potential of forward thinking.  Faced with an uninterrupted string of defections by the partner, the individual sees the advantage of sacrificing individual survival by cooperating early in the game, if the loner survival probability is relatively small ($a_0<a_0^{\prime\prime}$) and each additional sacrifice in individual survival increases pair survival ($bc>d^2$).  It is interesting that this advantage extends to $\lceil j^{\ast} \rceil = \lfloor i_C \rfloor$ which is not a function of the length of the game or of the partner's strategy as long as $i \geq \lceil i_C \rceil$ but only of the proximity to the end of the game.

\subsection{Global evolutionarily stable strategies} \label{sec:ESSglob}

From Section~\ref{sec:Global equilibrium}, we know that each isolated, local ESS of Section~\ref{sec:isolated} is globally stable.  If it also satisfies (\ref{eq:ess}), that is if $A(S_i;S_j) > A(S_j;S_j)$ for all $j \neq i$, then it is a global ESS.  This additional criterion means that, against a partner who adopts any alternative strategy, an individual who keeps the globally stable strategy does better than an individual who adopts the alternative strategy along with the partner.  This is clearly the case in Fig.~\ref{fig:3morewalks}C, where all vertical arrows end at the same globally stable state.  In fact, this criterion will always be met for alternative strategies with larger numbers of defections, since $j^\ast$ does not depend on the partner strategy.  However, it will not necessarily be met for alternative strategies with smaller numbers of defections, in particular when the candidate ESS is strategy $S_n$ as in Figure~\ref{fig:3morewalks}A.  Although the differences in payoff are not great for the parameters of Fig.~\ref{fig:3morewalks}A, for this example we may verify that $A(S_{20};S_0)=0.491 < A(S_{20};S_0)=0.496$.  We may conclude that a local ESS may be a global ESS but it does not have to be one. 

We could make $S_n$, or all-$D$, an ESS in all three examples of Fig.~\ref{fig:3morewalks} simply by making the game shorter: $n \leq 8$ in Fig.~\ref{fig:3morewalks}A, $n \leq 6$ in Fig.~\ref{fig:3morewalks}B, and $n \leq 4$ in Fig.~\ref{fig:3morewalks}C.  In addition, $S_n$ will be an ESS if $a_0 \geq a_0^\ast$, so that $J_{opt}$ does not exist.  The latter is a special case of defection always being favored (first, third and sixth possibilities in Table~\ref{tab:cases}) in which $S_n$ would be an ESS regardless of $n$.  All-$C$, or $S_0$, is never an ESS because defection is always favored in the final step of the game.  However, $S_1$ will be an ESS if $a_0$ is sufficiently small.  Finally, we may note that whereas Fig.~\ref{fig:3morewalks}C represents the tenth possibility in Table~\ref{tab:cases}, in which only a single equilibrium point is possible, an ESS for $i<n$ may also occur in the fifth possibility in Table~\ref{tab:cases}.  Simply changing $d$ from $0.949$ to $0.95$ in the example of Fig.~\ref{fig:3morewalks}C moves it from the tenth to the fifth possibility in Table~\ref{tab:cases}, by making $a_0^\prime > a_0^{\prime\prime}$, but results in virtually the same graph with $S_4$ as an ESS. 

\section{Discussion}

We have established some basic properties of strategy choice in iterated, two-player survival games, focusing especially on the case where each step is a Prisoner's Dilemma.  It would be of interest to investigate arbitrary strategies, including mixed strategies and reactive strategies, but for simplicity we have focused on pure, non-reactive strategies which switch from $C$ to $D$ at some step of the game.  We have denoted these by the number of end-of-game defections: $S_i$ means $C$ for $n-i$ steps then $D$ for $i$ steps, with $i \in [0,n]$.  Thus, the state space of strategies is an $(n+1) \times (n+1)$ matrix.   Our goal has been to understand how the payoff function $A(S_j;S_i)$, which is the survival probability of an individual with strategy $S_j$ whose partner has strategy $S_i$, depends on the parameters ($i,j,n,a,b,c,d,a_0$).  

Previous studies have addressed strategy choice in iterated survival games, but only under the assumption that an initial choice of a single-step strategy is maintained over the entire game. \citet{EshelAndWeinshall1988} modeled such constant, single-step strategies as probabilistic mixtures of $C$ and $D$ in the case that $n$ is geometrically distributed and $(a,b,c,d)$ in each step is randomly sampled from a distribution which assign non-zero probabilities to Harmony Games $(a \geq c,b \geq d)$ as well as to Prisoner's Dilemma's $(c>a>d>b)$; note this is our notation not theirs. \citet{EshelAndShaked2001} included the possibility of non-independence of players' survival in each step.  \citet{Garay2009} considered mixtures like those of \citet{EshelAndWeinshall1988} but in a game of fixed length and with constant single-step payoffs.  \citet{WakeleyAndNowak2019} studied the choice between two pure, single-step strategies in a fixed-length game.  

By considering the consequences of switching from $C$ to $D$ during the game in the case that each step is a canonical Prisoner's Dilemma ($c > a > d > b$, $a^2 > bc$) we found three critical values ($a_0^\ast,a_0^\prime,a_0^{\prime\prime}$) for the loner survival probability $a_0$ which establish broad patterns of incentives to cooperate or defect.  If $a_0 < a_0^\ast$, then an optimal number of defections $J_{opt}$ exists against a partner who never defects ($i=0$).  If $a_0 < a_0^\prime$, then a switch-point $i_D$ exists such that additional defection is favored for $i<i_D$ but disfavored for $i>i_D$.  If $a_0 < a_0^{\prime\prime}$, then a switch-point $i_C$ exists such that additional cooperation is favored for $i<i_C$ but disfavored for $i>i_C$.  These critical values are averages, each falling between an identical-pair survival probability and the corresponding individual survival probability: specifically between $a^2$ and $a$ in the case of $a_0^\ast$, and between $d^2$ and $d$ in the cases of $a_0^\prime$ and $a_0^{\prime\prime}$.  We have $a_0^\ast > a_0^\prime$, so the existence of $i_D$ guarantees the existence of $J_{opt}$ but not vice versa.  Further, depending on the parameters ($a,b,c,d$), $a_0^\prime$ may be either larger or smaller than $a_0^{\prime\prime}$, with important consequences for the structure of incentives.  

Extending the idea of $J_{opt}$ to other partner strategies, we found a single optimal response $j^\ast = \lfloor i_C \rfloor$ to any partner who defects more than $i_C$ times, and a series of optimal responses $i + k(i)$, beginning at $J_{opt}$ for $i=0$ and ending at $\lceil i_D \rceil$ for $i = \lfloor i_D \rfloor$, to a partner who defects fewer than $i_D$ times.  When $i_D$ exists, a stretch of equilibria may exist, composed of stable strategies for which there is no incentive either to cooperate more or to defect more.  The stretch extends from $\lceil i_D \rceil$ to $\lfloor i_C \rfloor$ if $i_C$ exists and $\lceil i_D \rceil \leq \lfloor i_C \rfloor$, or to $n$ if $i_C$ does not exist or if $\lceil i_D \rceil \leq n \leq \lfloor i_C \rfloor$.  Alternatively, when $i_C$ exists, a stretch of disequilibria may exist, composed of unstable strategies for which there is incentive both to cooperate more and to defect more.  These stretches extend from $\lceil i_C \rceil$ to $\lfloor i_D \rfloor$ if $i_D$ exists and $\lceil i_C \rceil \leq \lfloor i_D \rfloor$, or to $n$ if $i_D$ does not exist or if $\lceil i_C \rceil \leq n \leq \lfloor i_D \rfloor$.  When neither $i_D$ nor $i_C$ exist or when $n<i_D,i_C$, additional defection is favored such that the single best strategy is $S_n$.  Other special cases occur; Table~\ref{tab:cases} lists all possibilities.  

Two more general features of our model are notable.  First, strategy choice depends explicitly on the number of steps left in the game, but only incidentally on its length.  The parameter $n$ of course affects the magnitude of the overall payoffs.  But it is possible to ignore $n$ in the describing the properties of $J_{opt}$, $i_D$, $i_C$, etc., and only later bring $n$ in as an upper bound to specify whether some of these quantities might be irrelevant in a given game.  Second, $J_{opt}$, $i_D$ and $i_C$ are all increasing functions of $a_0$.  They are J-shaped, staring near zero for small $a_0$ and diverging as $a_0$ approaches the corresponding critical value.  If the loner survival probability $a_0$ is small, the incentive to defect only arises near the end of the game.  But if $a_0$ is close to one, the incentive to cooperate in an iterated survival game disappears completely. 

Using the notion of a best-response walk, we showed that stretches of both equilibria and disequilibria are approached from above and below.  Stretches of disequilibria often lead to cycles between two or more strategies.  Walks approaching stretches of equilibria hit the endpoints but do not enter the interior.  We analyzed equilibria from the standpoint of evolutionary stability, and showed that equilibrium strategies are not protected against invasion by other equilibrium strategies with fewer defections.  For a strategy to be an ESS it must be the only equilibrium strategy. However, the converse is not true. 

The natural scale of survivability facilitates the investigation of all possible survival games.  We have delineated the possibilities for iterated survival games in which individuals may switch from $C$ to $D$ once during the game, under the assumption that the single-step game is a Prisoner's Dilemma.  In many cases, the essential structure of the Prisoner's Dilemma is undermined upon iteration.  In closing, we explore the parameter space to gauge how broadly cooperation may be supported in these games.  Table~\ref{tab:fivestructures} shows the fractions of times that five qualitatively different incentive structures for cooperation occurred when survival probabilities ($a,b,c,d,a_0$) were sampled uniformly at random under two different models.  

\begin{table}[h]
{\centering
 \begin{tabular}{lcc}
  \hline 
  & \multicolumn{2}{c}{Models for Random Sampling} \\ \cline{2-3}
                      & $a,b,c,d \in (0,1)$ & $a,b,c,d \in (0.9,1)$\\[-4pt]
  Incentive Structure & $a_0 \in (0,1)$     & $a_0 \in (0.7,1)$\\ 
  \hline 
  1. defection always favored           & $66.59$\%& $22.95$\% \\ 
  2. unbounded stretch of equilibria    & $24.22$\%& $30.20$\% \\ 
  3. bounded stretch of equilibria      & $6.70$\%& $25.69$\% \\ 
  4. unbounded stretch of disequilibria & $1.04$\%& $2.31$\% \\ 
  5. bounded stretch of disequilibria   & $1.45$\%& $18.85$\% \\ 
  \hline 
 \end{tabular}
\caption{Outcomes for one million parameter sets sampled uniformly at random according to two different models, in which the single-step game is a Prisoner's Dilemma ($c>a>d>b$ and $a^2>bc$).} \label{tab:fivestructures}
}
\end{table}

Specifically, we binned the ten possibilities in Table~\ref{tab:cases} into five types of incentive structures.  Type 1 includes the first, third and sixth possibilities.  These are all cases in which additional defection is favored (and additional cooperation disfavored) against all possible partner strategies.  In other words, neither $i_D$ nor $i_C$ exists.  Figure~\ref{fig:3morewalks}A shows an example (in which $J_{opt}$ does exist).  Type 2 includes the second and fourth possibilities, in which $i_D$ exists but $i_C$ doesn't, producing a stretch of equilibria which begins at $\lceil i_D \rceil$ and has no upper bound except $n$. Figure~\ref{fig:3morewalks}B shows an example.  Type 3 includes the fifth and tenth possibilities, in which both $i_D$ and $i_C$ exist and there a stretch of equilibria from $\lceil i_D \rceil$ to $\lfloor i_C \rfloor$.  Figure~\ref{fig:standard walk}AC and Fig.~\ref{fig:3morewalks}C show examples. Type 4 includes just the seventh possibility, in which $i_C$ exists but $i_D$ doesn't, producing a stretch of disequilibria which begins at $\lceil i_C \rceil$ and has no upper bound except $n$.  We have not depicted this case, but note that it leads to multi-state cycles in best-response walks.  Type 5 includes the eighth and ninth possibilities, in which both $i_D$ and $i_C$ exist and either there is a stretch of disequilibria from $\lceil i_C \rceil$ to $\lfloor i_D \rfloor$ or there are no integers between $i_D$ and $i_C$. Figure~\ref{fig:standard walk}BD shows an example of the former.  The latter leads to cycles between two adjacent states (not shown). 

We considered two different ranges of survival probabilities as models for random sampling.  The first model samples uniformly over the entire parameter space.  This covers all possible iterated survival games, including many cases when there is no advantage at all to having a partner ($a_0 > a,b,c,d$).  The second model samples over two narrower ranges, $0.9$ to $1$ for $a$, $b$, $c$ and $d$, and $0.7$ to $1$ for $a_0$.  This captures the range of examples we have presented in this work.  Sampling $a,b,c,d>0.9$ represents games which are, arguably, relatively mild in a single step but may become very harsh upon iteration.  The resulting single-step pairwise survival probabilities, $a^2$, $bc$ and $d^2$, will all be greater than 0.8.  Sampling $a_0 > 0.7$ then covers a range of models with relatively bleak prospects for loners, which should favor cooperation, but also allows that $a_0$ might be comparable in magnitude to, or even greater than $a$, $b$, $c$ and $d$.  

We took one million random samples of parameters for each model.  We assigned parameter labels such that $c>a>d>b$, then excluded samples which did not satisfy $a^2 > bc$.  This excluded about $10$\% of samples in the first model and about $24$\% in the second model.  We checked the remaining samples against the criteria in Table~\ref{tab:cases}, then binned them into the five qualitatively different incentive structures and computed the percentages of samples falling under each type of structure.

Table~\ref{tab:fivestructures} illustrates the ways in which cooperation can be favored in iterated survival games, in terms of fractions of the parameter space.  For the first model, which samples over all possible parameters $a,b,c,d,a_0 \in (0,1)$, about two-thirds of parameter sets yield games in which additional defection is favored against any partner strategy.  Most of the other one-third of the parameter space corresponds to games with an unbounded stretch of equilibria.  Games with stretches of disequilibria are rare.  Given that the single-step game is a Prisoner's Dilemma, cooperation may be said to be favored whenever equilibria or disequilibria exists, at least in the sense of there being checks on the number of end-game defections.  

In fact, due to the shapes of $i_D$ and $i_C$ as functions of $a_0$, which remain relatively small until diverging sharply as $a_0$ approaches $a_0^\prime$ and $a_0^{\prime\prime}$, there are essentially two kinds of games.  On the one hand, if $a_0 \geq a_0^\prime,a_0^{\prime\prime}$, defection is clearly favored.  On the other hand, if $a_0<a_0^\prime$ or $a_0<a_0^{\prime\prime}$, there are strong checks on defection.  Across all cases in which $i_D$ or $i_C$ exists in Table~\ref{tab:fivestructures}, the median $i_D$ was $0.3$ and the $90$th percentile $i_D$ was $2.0$.  The median $i_C$ was $1.6$ and the $90$th percentile $i_C$ was $4.6$.  We might also point out that in the case of an unbounded stretch of equilibria, the results in Section~\ref{sec:diagonal} show that none of the equilibrium strategies are protected against invasion by strategies with smaller numbers of defections. 

As expected for the second model, with $a,b,c,d \in (0.9,1)$ and $a_0 \in (0.7,1)$, cooperation is favored over a larger fraction of the sampled parameter space.  Defection is favored in less than one-quarter of games.  Bounded stretches of equilibria or disequilibria are more frequent.  Unbounded stretches of disequilibria remain rare, which makes sense because this requires that $a_0$ falls between $a_0^\prime$ and $a_0^{\prime\prime}$.  Even over the restricted parameter space of this sampling model, there is a dramatic difference between games in which defection is always favored and games in which cooperation is favored in the sense of there being checks on the number of end-game defections.  Here, across all cases in which $i_D$ or $i_C$ exists, the median $i_D$ was $2.4$ and the $90$th percentile $i_D$ was $16.8$; the median $i_C$ was $4.5$ and the $90$th percentile $i_D$ was $22.5$.  Overall, using this sampling model or the previous one to frame the results of Sections~\ref{sec:fullcoop} through~\ref{sec:global}, we find surprisingly strong support for cooperation in iterated survival games, mediated by the loner survival probability. 

\section*{Acknowledgements}

Olivier Salagnac was supported by a fellowship from the \'{E}cole Normale Sup\'{e}rieure.  The question about backward induction which motivates Section~\ref{sec:fullcoop} was posed by Martin Nowak, who also provided helpful advice about the criteria for Nash equilibria and ESSs, (\ref{eq:nash}) and (\ref{eq:ess}).  


\bibliography{refs}

\end{document}